\documentclass[a4paper,usenatbib]{mnras}

\makeatletter

\newcommand{\Rmnum}[1]{\expandafter\@slowromancap\romannumeral #1@}

\newcommand{\gsim}{\lower0.6ex\vbox{\hbox{$\buildrel{\textstyle >}\over{\sim}\ $}}}

\makeatother

\usepackage{newtxtext,newtxmath}
\usepackage{graphicx}	
\usepackage{amsmath}	
\usepackage{subcaption}
\usepackage{booktabs}
\usepackage{float}
\usepackage[dvipsnames,svgnames]{xcolor}
\usepackage{color}

\def\hmpc{h^{-1}{\rm Mpc}}

\def\hmsun{{h^{-1} M_{\odot}}}
\def\msun{\, M_{\odot}}

\title[Super-resolution simulations]
{AI-assisted super-resolution cosmological simulations III: Time evolution}

\author[X. Zhang et al.]
{Xiaowen Zhang$^{1,2}$\thanks{Email:xiaowen4@andrew.cmu.edu}, Patrick Lachance$^{1,2}$, Yueying Ni$^{3}$, Yin Li$^{4}$,
Rupert A.~C. Croft$^{1,2}$,
\newauthor Tiziana Di Matteo$^{1,2}$, Simeon Bird$^{5}$, Yu Feng$^{6}$ \\
$^1$ McWilliams Center for Cosmology, Department of Physics, Carnegie Mellon University, Pittsburgh, PA 15213 \\
$^2$ NSF AI Planning Institute for Physics of the Future, 
Carnegie   Mellon  University, Pittsburgh, PA 15213, USA \\
$^3$ Harvard-Smithsonian Center for Astrophysics, 60 Garden Street, Cambridge, MA 02138, USA \\
$^4$ Department of Mathematics and Theory, Peng Cheng Laboratory, Shenzhen, Guangdong
518066, China \\
$^5$ Department of Physics and Astronomy, University of California Riverside,
900 University Ave, Riverside, CA 92521 \\
$^6$ Berkeley Center for Cosmological Physics and Department of Physics, University of California, Berkeley, CA 94720, USA \\
}

\date{Accepted XXX. Received YYY; in original form ZZZ}

\pubyear{2022}
\begin{document}
\maketitle

\begin{abstract}
In this work, we extend our recently developed super-resolution (SR) model for cosmological simulations to produce fully time-consistent
evolving 
representations of the particle phase-space distribution.
We employ a style-based constrained generative adversarial network (Style-GAN) where the changing cosmic time is an input style parameter to the network. 
The matter power spectrum and halo mass function agree well
with results from high-resolution N-body simulations over the full trained redshift range ($10 \le z \le 0$).
Furthermore, we assess the temporal consistency of our SR model by constructing halo merger trees. 
We examine progenitors, descendants and mass growth along the tree branches. All statistical indicators demonstrate the ability of our SR model to generate satisfactory high-resolution simulations based on low-resolution inputs.
\end{abstract}


\section{Introduction}
\label{section1:introduction}

As future cosmological surveys aim to cover the sky more broadly and deeply, there is a growing demand for cosmological simulations with larger sizes and finer resolutions \cite[see e.g.][for a review]{Vogelsberger}. These simulations are essential for making predictions from theoretical models and for comparison with galaxy surveys. Cosmological surveys such as Euclid \citep{euclid} and the Rubin Telescope Legacy Survey of Space and Time (LSST) \citep{LSST} will require a large number of high-resolution simulations to gain insight into cosmological models and how they apply to our Universe. 

Cosmological N-body simulations are powerful numericals tool for solving the non-linear evolution of cosmic structure formation. While they are computing-intensive, high-resolution dark matter-only simulations using N-body codes, such as \texttt{MP-Gadget}\footnote{\url{https://github.com/MP-Gadget/MP-Gadget}}, can follow the small-scale evolution of galaxy halos and so allow the creation of detailed merger trees. Even with high-performance computing resources, these numerical models require a significant amount of storage and computation due to the complex and non-linear dynamics involved in gravitational structure formation, which is a very challenging task. This limitation often requires researchers to choose between maximizing resolution or volume. These different needs are illustrated by the
range of dark-matter and hydrodynamical simulations carried out in recent years, from the small volume high resolution FIRE suite \citep{fire} and to the ABACUS summit\citep{abacus} optimized for large-scale structure. 

In recent years, Machine learning (ML) \citep{Yue2016} has become a promising tool in physics, capable of solving non-linear problems and reducing computation times.  ML is now being extensively applied to cosmology \citep{MLincosmo}, addressing problems that were previously difficult to solve. Deep Learning (DL), the use of neural networks \cite{MLincosmo}, has found applications in many different aspect in cosmological simulations. For example, using DL,  non-linear structures can be derived directly from cosmological initial conditions \citep{He_2019}, and mock halo catalogs can be inferred from density fields \citep{berger2019, Bernardini2020}. Other  works use Generative Adversarial Networks and denoising diffusion model(GANs, \citep{goodfellow2014GAN, denoising}), learning from  2D images of cosmic webs \citep{Rodrguez2018} to generate synthetic versions. Matter density fields in 3D have
been generated also \citep{Perraudin2019}. 
Machine learning models have been developed to predict different baryonic properties from dark matter-only simulations, including the distribution of galaxies, thermal Sunyaev-Zeldovich (tSZ) effect, 21 cm emission from neutral hydrogen, stellar maps, and various gas properties. Examples of these studies are some that have used machine learning to predict the galaxy distribution \citep{Modi2018, zhang2019}, others the tSZ effect \citep{Troster2019}, the 21 cm emission distribution \citep{wadekar2020hinet}, and gas properties such as stellar maps \citep{Dai2021}.
CAMELS (Cosmology and Astrophysics with MachinE Learning Simulations, \cite{}), is a large project involving  a training set of over 4000 hydrodynamical simulations run with different hydrodynamic solvers and subgrid models for galaxy formation. A prime purpose of this set is to investigate the interplay of baryonic effects and cosmology, using machine learning techniques.

Restoring high-resolution information from low-resolution data ("super-resolution") is a challenging task, as there are many possible high-resolution solutions to a given low-resolution input.
Deep learning (DL) techniques have shown promise in addressing this issue(see \citep{srreview} for a review), with one of the most promising models being Generative Adversarial Networks (GANs) \citep{goodfellow2014GAN}. GANs are generative models that train both a generator and discriminator simultaneously, through adversarial competition. A Super-resolution GAN (SRGAN) \citep{SRgan}  uses a generator to learn from the distribution of low-resolution (LR) fields and produce a high-resolution (HR) distribution, while the discriminator estimates the probability of whether the field is from HR or SR. During training, both the generator and discriminator improve their performance, and eventually, they reach a local Nash equilibrium, where the discriminator becomes unable to differentiate between HR and SR fields.
Several studies have explored the application of super-resolution (SR) techniques \cite{} to  the spatial or mass resolution of cosmological simulations. \citep{KodiRamanah2020}, for instance, developed a SR network that maps the density fields of low-resolution (LR) cosmological simulations to those of high-resolution (HR) ones. This approach has the potential to significantly reduce computational costs while enabling the exploration of more extensive parameter spaces in cosmological simulations.

In our previous work \citep{AI1,AI2} (hereafter Paper I and Paper II), we showed using cosmological
simulations that the output of a GAN using a low resolution input can preserve the input large scale structure and also create small scale structures. These satisfy statistical properties of the high resolution simulation such as the matter power spectrum and halo mass function, as well as the visual aspects of structures. The SR models are capable of generating the same number of particles as HR but four orders of magnitude faster than a highly parallel N-body code. 

However, the current SR models are restricted to a single redshift due to their training on LR and HR simulations at that specific redshift. To generate SR fields at different redshifts, the DL model used in Papers I and II need to be retrained with new sets of LR and HR simulations. In this paper (paper III) we present new work
that removes this restriction, allowing SR simulations to be generated at any redshift following a single training episode. To do this, we use the capabilities of the NN architecture
StyleGAN \citep{stylegan2}, which is capable of mapping random noise to a latent vector that represents the style parameters of an image. These parameters can be manipulated to generate diverse outputs with distinct features such as shape and color. Incorporating a style-based generator and discriminator into the SR model will facilitate the generation of SR fields at different redshifts by allowing the neural network to generalize its outputs. 
In this work all simulations in the training and test sets will have the same cosmological parameters,  and we will be using the cosmic scale factor $a$ as our style parameter, the natural choice. 
This style parameter as an input alongside the LR simulation will allow the NN to learn about extra information about 
the time evolution of the dark matter distribution compared to our
previous architecture. This approach, using a style parameter, can also be used to encode the cosmology dependence in NN and enabling it to effectively interpolate the results of structure formation across cosmologies \citep[see recent work by][]{jamieson2022field}. In our work, by sampling over scale factor, we create different simulations with different initial random seeds and save snapshots at certain scale factor, compiling all snapshots into a single training set. The advances in this paper are the following:

\begin{itemize}
    \item We incorporate the scale factor as a style parameter in addition to the input 6D phase space (displacement and velocity).
    \item To evaluate the accuracy of time evolution afforded by the new SR-GAN, we compute dark matter halo merger trees from
    the SR outputs. We utilize a time-consistent LR-HR pair as our test set. By examining various statistics, such as the length of merger trees and branching ratios, we can determine the authenticity of the generated merger trees.
\end{itemize}
The paper is structured as follows: In Section 2, we review  our neural network architecture, the training and test sets, as well as the training process. In Section 3, we present our results,  discussing statistics related to the phase space of the generated field in Sections 3.1 and 3.2, while Section 3.3 focuses on merger trees and related statistics. In Section 4, we discuss our findings, and conclude in Section 5.

\section{Method}
\label{section2:Method}

\begin{figure*}
\includegraphics[width=\textwidth]{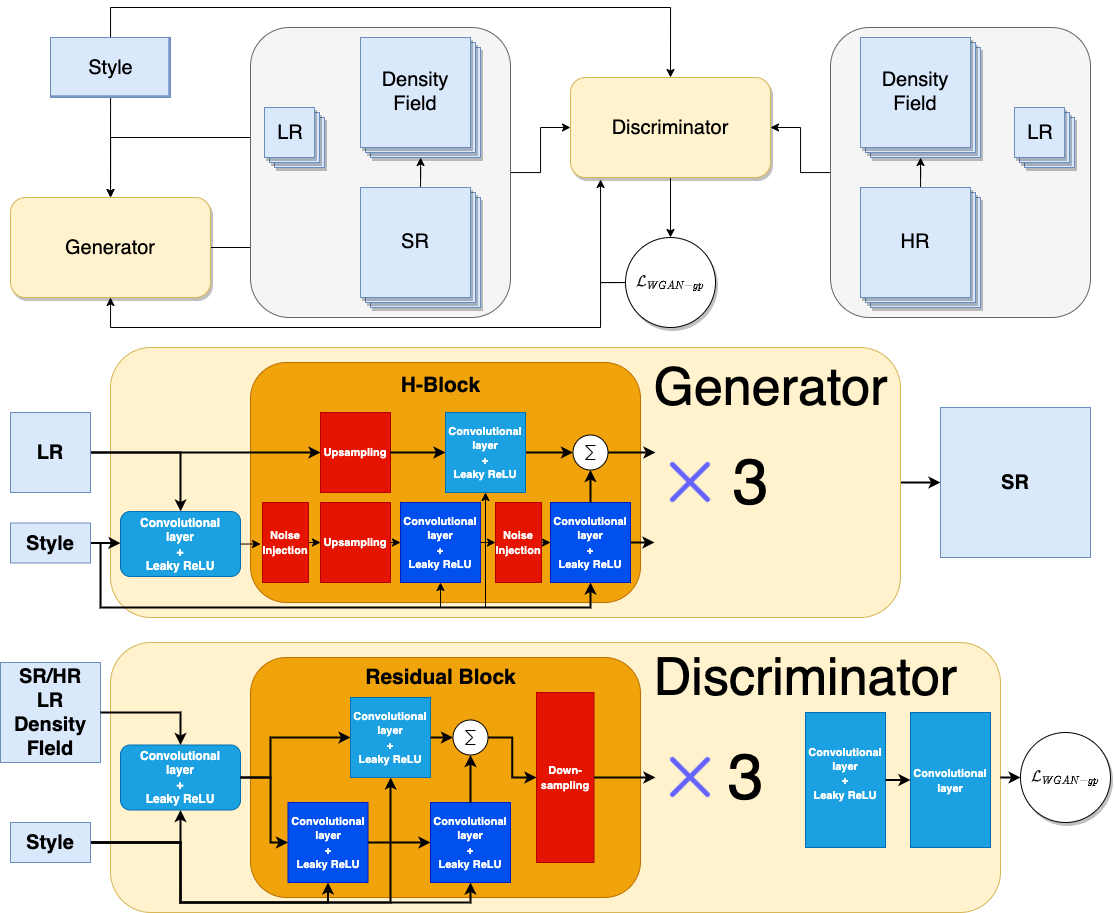}
\caption{
\textit{\textbf{Upper panel:}}
Schematic plot of our GAN training process. 
\textit{\textbf{Middle panel:}}
The architecture of the generator. Dark blue blocks denote size 3 kernel size, and light blue denotes size 1 kernel convolutional layer. Other operations like upsampling and noise injection blocks are colored in red.
\textit{\textbf{Lower panel:}}
The architecture of the discriminator.
}
\label{fig:MLStructure}
\end{figure*}

\subsection{$N$-body simulation}

The $N$-body simulation is a widely used technique for understanding the non-linear development of cosmic structures.
 Given a fixed cosmological matter (as prescribed by our standard cosmological model \cite{hinshaw13}), this method involves dividing the mass distribution into $N$ equal-mass particles, which are positioned initially using a Gaussian random field with a specific power spectrum (determined by cosmological parameters). To predict the velocity and displacement of each particle, Poisson's equation is solved numerically for dark matter-only simulations. The value of $N$ affects the mass resolution of the simulation, with higher values making it more detailed but also slower due to the need to increased numbers of force calculations and  greater precision. Despite the use of high-performance computing resources, producing high-resolution $N$-body simulations is a costly process because the computational complexity for the most efficient methods increases with $\mathcal{O} (N \log{}N)$ . We run simulations using the N-body code \texttt{MP-Gadget}\footnote{\url{https://github.com/MP-Gadget/MP-Gadget}}, a highly parallel program that solves the gravitational force using the TreePM method \citep{bagla02}, with Fast Fourier computation of long-range forces and a hierarchical octree algorithm for short-range forces.

We conduct LR and HR simulation tasks in the Lagrangian description with particles originally positioned on a uniform grid. The displacement field is:
\begin{align}
    \mathbf{d}_i = \mathbf{x}_i - \mathbf{q}_i
\end{align}
where $\mathbf{x}_i$ is the current position of a particle and $\mathbf{q}_i$ is its initial position. Similarly, the velocity field of each particle is :
\begin{align}
    \dot{\mathbf{d}_i} = \dot{\mathbf{x}_i} - \dot{\mathbf{q}_i}
\end{align}

There are several advantages to using Lagrangian description of the cosmic density rather than an Eulerian description:
\begin{itemize}
    \item The total mass of the output field is conserved because mass, not space, is discretized.
    \item In the Lagrangian description, resolving small structures is improved by learning the displacement of tracer particles instead of the density field.
    \item As we shall show, the tracer particles can be traced through time, making the SR field statistically identical to the real output from the $N$-body simulation.
\end{itemize}

Consequently, the Lagrangian description provides a more accurate description of fields with a larger dynamic range compared to the Eulerian description with the same grid size. In the Eulerian description, the simulation field must be mapped onto a uniform grid with limited resolution, while in the Lagrangian description, much smaller scales can be more accurately resolved.

\subsection{Generative adversarial networks}

Generative adversarial networks consist of two neural networks that compete with each other and train simultaneously. The generator $G$ generates fake samples, and discriminator $D$ attempts to find the difference between real and fake samples and gives feedback to generator $G$. During the training, generator $G$ will be able to generate more realistic samples while discriminator $D$ improves by learning the detailed differences between real and fake data. However, optimizing GANs can be very difficult because it involves finding the Nash equilibrium between the generator and discriminator which is a non-convex problem and can have multiple local minima. Additionally, a GAN could also suffer from mode collapse and gradient diminishing problems (see 
\cite{}). In this paper, we follow the StyleGAN2 architecture and minimize the Wasserstein \citep{wgan} distance between real and fake samples, which can avoid these difficulties. 
We also use gradient penalty(WGAN-GP) \citep{wgan-gp} to achieve a soft version of the Lipschitz constraint instead of hard clipping as in the original WGAN.
The WGAN-GP loss function is:
\begin{multline}
  L_\mathrm{WGAN-gp} = \mathrm{E}_{l, z} [D(l, G(l, s, z))]
  - \mathrm{E}_{l,h} [D(l, s, h)] \\
  + \lambda \; \mathrm{E}_{l, h} \bigl[\bigl( \|\nabla_i D(l, s, i)\|_2 - 1 \bigr)^2\bigr].
  \label{L_GAN}
\end{multline}
The Wasserstein distance is the first line in the objective function, while the second line represents the gradient penalty. A random sample, denoted as $i$, is drawn uniformly along the lines connecting pairs of real fields (h) and fake fields ($G(l, s, z)$). The hyperparameter $\lambda$ is used to control the magnitude of the gradient penalty. The discriminator $D$ is optimized by maximizing all three terms, whereas the generator $G$ optimizes only the first term. We use the most common approach, and apply the penalty every 16 batches.

\subsubsection{Training process}

The goal of our SR task is to increase the particle number $N$, compared to an input LR realization by a factor of 512 at any redshift required. Success in this approach will significantly reduce the computational time required to run N-body simulations with comparable accuracy in many aspects to
HR simulations.

The training process is illustrated in the upper panel of Figure \ref{fig:MLStructure}. The displacement and velocities of particles in both the LR and HR/SR fields are concatenated to form a 6-channel input field. The corresponding scale factor is stored in a size-1 array and input into the neural network along with the concatenated field. Due to GPU memory constraints, the LR and HR/SR particles are cropped into cubic chunks with added padding margins and periodic boundaries.

In the GAN model, the generator $G$ transforms the LR input field $l$ and scale factor $s$ into SR displacements and velocities, $G(l, s)$, with 512 times the LR resolution. The relative distance between the SR and HR fields is measured using the mean-square error (MSE) as the loss function, which is not used in the training process. Both the LR and HR fields are concatenated and fed into the discriminator $D$, which evaluates the authenticity of the combined fields. The size of the LR field is matched to the SR and HR fields using tri-linear interpolation.

Additionally, the Eulerian space density field \citep{shi2016real} is also concatenated to the inputs of the discriminator, which is calculated from the Lagrangian representation of the displacement field using a differentiable Cloud-in-Cell operation. This provides the discriminator with the ability to visualize the structure directly in Eulerian space. This is crucial for accurately predicting visually distinct features and precise statistics of small structures.

\subsubsection{Details of the architecture}
\label{sub:arch}

Our model is based on the architecture presented in \citep{AI2} and incorporates elements from the StyleGAN2 structure. Figure \ref{fig:MLStructure} (middle panel) illustrates the generator's architecture. The generator consists of multiple H-blocks, each composed of two branches: the projection branch and the convolutional branch. The projection branch starts by up-sampling the input field by a factor of 2 using tri-linear interpolation and then goes through a 1-kernel size convolutional block followed by a Leaky ReLU activation function. The convolutional branch receives the output from a 1-kernel size convolutional block. The injection of random noise before and after the up-sampling process is crucial as it introduces stochasticity and generates new features that are not present in the input field. As we explain later, keeping this noise input 
identical for each realization of the cosmological density field, irrespective of cosmic time is crucial to generation of time-consistent evolution of cosmic structures in the SR simulation.
The output of the H-block is obtained by summing the outputs of both branches. All activation functions used are Leaky ReLU with a negative slope of 0.2. By stacking 3 H-blocks, the output field is up-sampled by a factor of 8 along each edge, resulting in 512 additional particles. The scale factor, serving as the style parameter, is fed into a single linear layer before being used to control the weights of each convolutional layer.

The discriminator network architecture is illustrated in the lower panel of Figure \ref{fig:MLStructure}. Class of fully convolutional discriminator like this is known as patchGAN \citep{isola2017image} It is designed as an inverse of the generator, and it comprises three residual blocks \citep{resnet}. Each residual block comprises two branches: the "skip" branch, consisting of a single 1-kernel size convolutional block, and the convolutional branch, which is composed of two 3-kernel size convolutional blocks, each followed by a leaky ReLU activation function. The output from the two branches is then summed and undergoes a down-sampling process by a factor of 2. The final output of the discriminator is a single channel critic, which is averaged to evaluate the Wasserstein distance.

\subsection{Training set and test set}

Our training dataset is comprised of 8 dark matter only simulations, each with low-resolution (LR) and high-resolution (HR) pairs. Each simulation contains 100 snapshots that uniformly sample logarithmic scale factor. The mean spatial separation of the dark matter particles is used to determine the gravitational softening length, which is set to $1/30$ of this value. The cosmological parameters for all simulations are based on the WMAP9 cosmology \citep{hinshaw13}, with matter density $\Omega _{\rm m} = 0.2814$, dark energy density $\Omega_{\Lambda} = 0.7186$, baryon density $\Omega {\rm b} = 0.0464$, power spectrum normalization $\sigma_{8} = 0.82$, spectral index $n_{s} = 0.971$, and Hubble parameter $h = 0.697$.


All the simulations have a box size of $(100,\hmpc)^3$ and are comprised of $64^3$ low-resolution (LR) and $512^3$ high-resolution (HR) dark matter particles. The LR particles have a mass resolution of $m_{\mathrm{DM}} = 2.98 \times 10^{11} \msun/h$ while the HR particles have a mass resolution of $m_{\mathrm{DM}} = 5.8 \times 10^{8} \msun/h$, which is $1/512$ of the LR mass resolution. Each of the eight simulations has a unique random seed for its initial conditions, resulting in distinct initial conditions for each simulation.

In addition, we also create 3 sets of dark-matter-only LR-HR simulation pairs, which share the same box size and cosmological parameters as the training set, but with a different random seed for the initial conditions. The test set consists of 95 LR-HR snapshot pairs, sampled uniformly in scale factor space from redshift of 5 to 0. Our evaluation of the model's performance is based on the test set, and the statistical results are compared against those from the corresponding LR and HR simulations. In addition, we generate 10 different test sets consist solely of integer redshifts(z=10, 7, 4, 3, 2, 1, 0). It is important to note that none of the integer redshifts are included in the training set, in order to evaluate the extrapolation capability of the SR model. 

\section{Results}
\label{section3:Result}


\begin{figure*}
\centering
  \includegraphics[height=0.99\textheight]{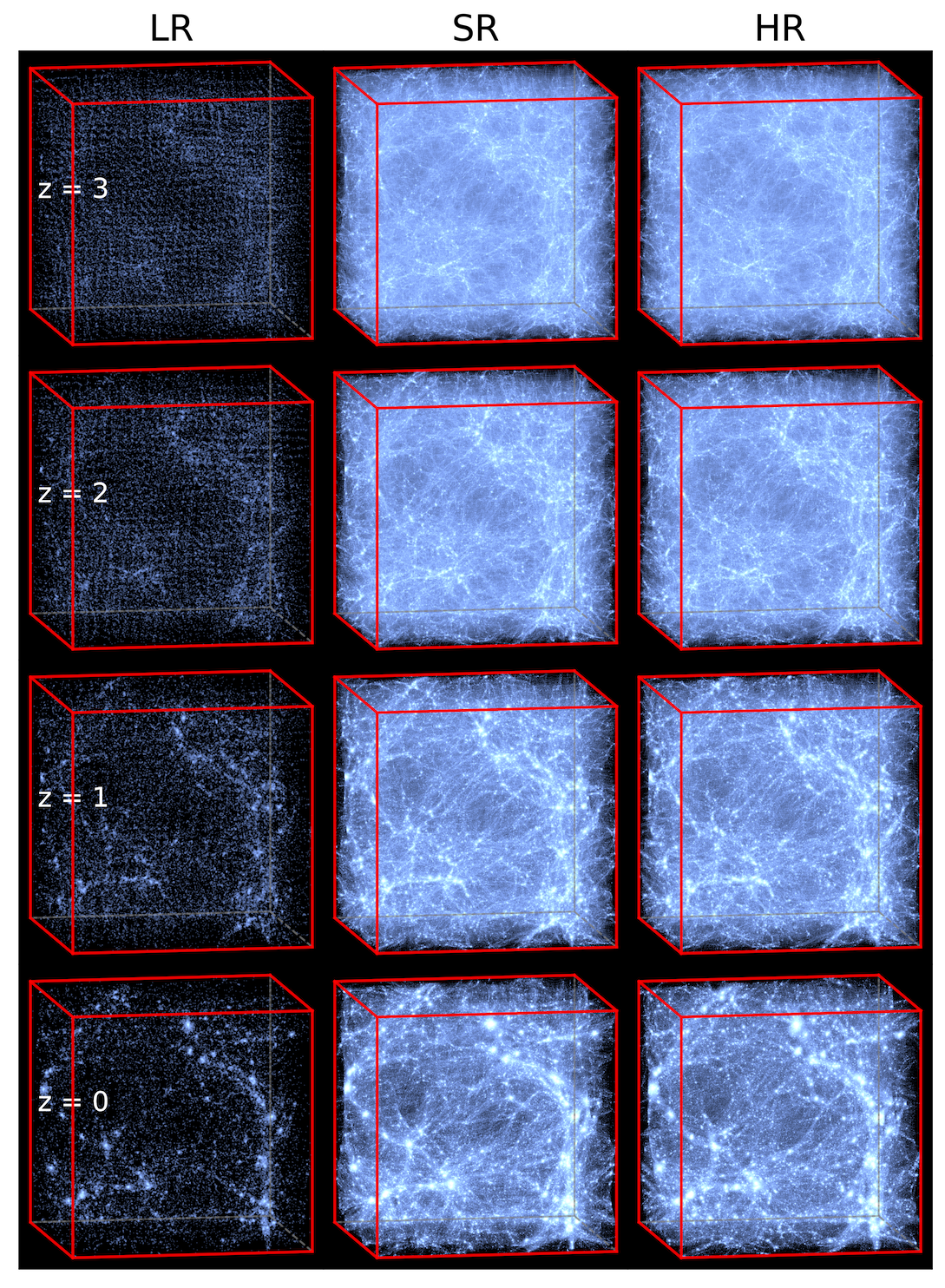}
  \caption{3D visualization of dark matter density field of whole $100 \hmpc$ simulation box and comparison to LR simulations at $z = 3, 2, 1, 0$.}
  \label{fig:3d-visual_wholebox}
\end{figure*}

\begin{figure*}
\centering
  \includegraphics[height=0.96\textheight]{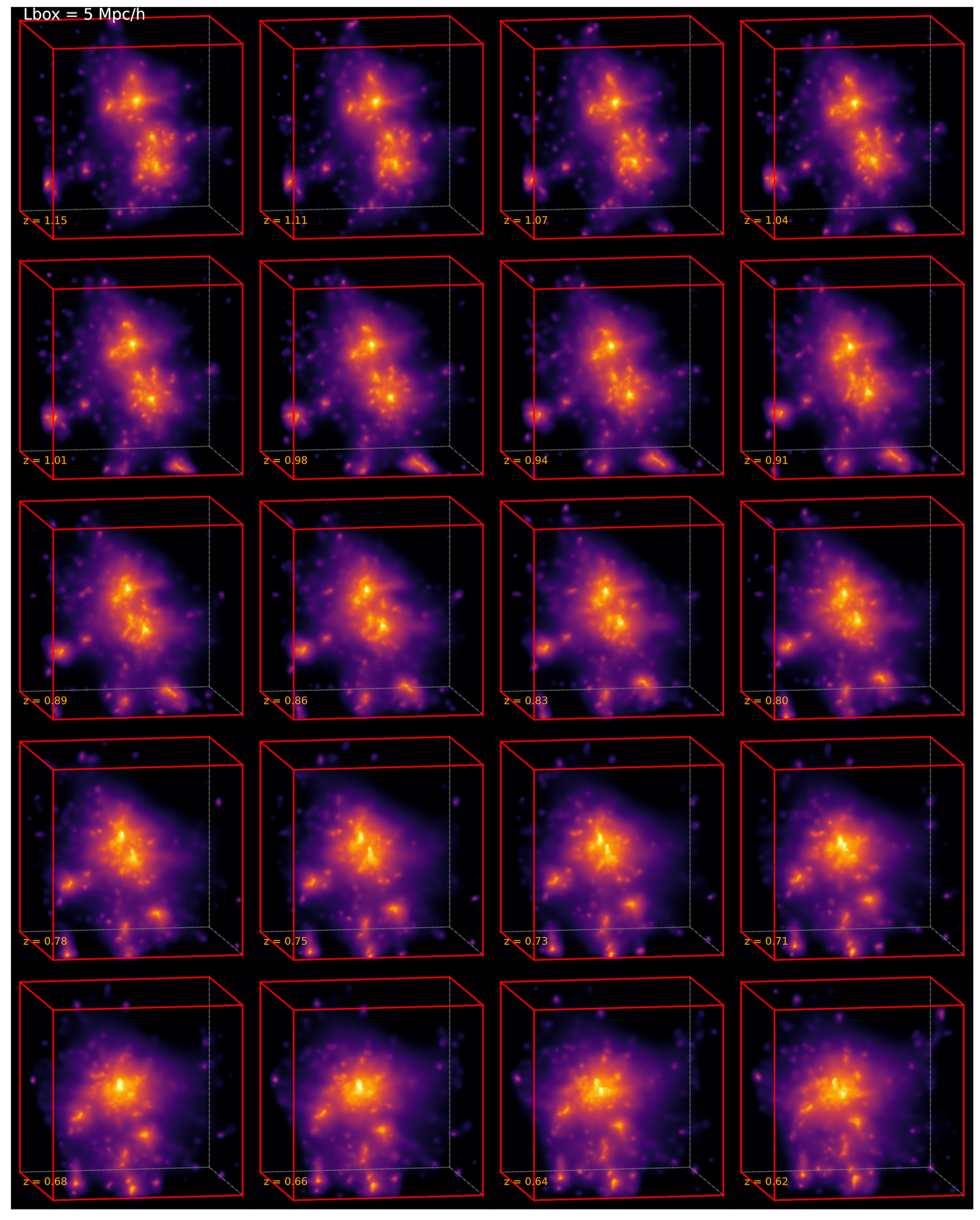}
  \caption{Time-evolving 3D projection of the matter density field with boxsize $ 5 \hmpc$, with time progressing from left to right and top to bottom, spanning the redshift range of $z = 1.1$ to $z = 0.6$}
  \label{fig:time_seq}
\end{figure*}

\begin{figure*}
    \centering
    \includegraphics[width=1.0\textwidth]{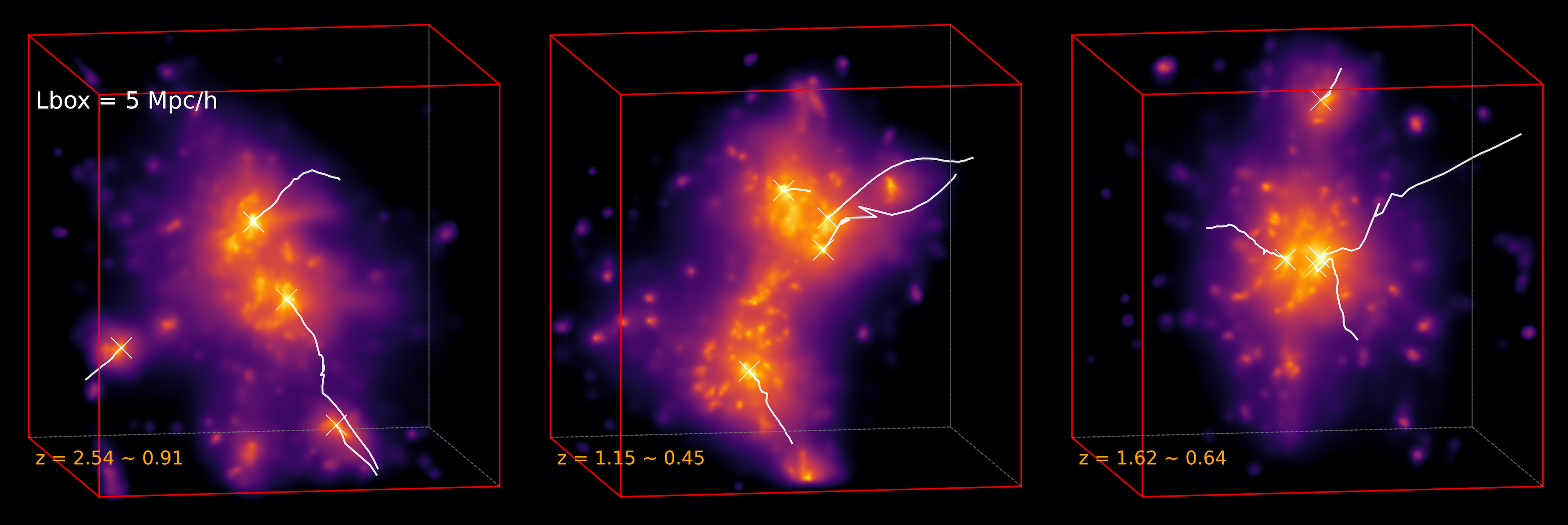}
    \caption{The halo trajectory plot. Each red box has the same side length $5 \hmpc$ from SR field and redshift range are labeled at lower left corner of each box. The top four most massive halos' trajectories are overlapped onto the 3D density field, X mark indicate the halo position at the last redshift.}
    \label{fig:halo_traj}
\end{figure*}

\label{subsection:visual}
We first evaluate the visual representation of high-resolution fields derived from $N$-body simulations and super-resolution fields from our hybrid Neural Network $N$-body calculations. The three-dimensional visualizations of dark matter density plots are produced by projecting all the particles within the entire $100 \hmpc$ simulation volume. 
Figure \ref{fig:3d-visual_wholebox} presents the 3D density visualizations at redshift 3, 2, 1, and 0. The images are rendered using the gaepsi2 code. The first column of the images displays the 3D matter density field of the LR simulations and are utilized to generate SR simulations (second column) by our NN , while the third column shows the corresponding HR field from the test set.

The image demonstrates that the large-scale structure is successfully captured by the NN, as we expect, given that it is conditioned on a low-resolution input. The small-scale structures generated by the NN are visually comparable to those from HR simulations, and their differences can be controlled by altering the random seed used in the noise injection to the NN. 

The visualization of the matter density field demonstrates that the SR field and HR simulation are essentially indistinguishable on large scales. On small scales, both fields are visually authentic, given the LR input field. The SR field accurately predicts the appearances of high-density regions both on large scale and small scales. These regions contain  most of the dark matter halos, and this motivates us to compare the halo abundance using other statistical methods.

In addition to the integer redshift test set, we generate a time evolution series using our time evolution test set and represent the 3D dark matter density in the form of a time evolution plot. The center of this sub-box is located at the position of the same Friends-of-Friends (FoF) halo with halo mass $10^{14} \hmsun $ . A movie version of time evolution is available in our github repository (see in Data Availability chapter).
In Figure \ref{fig:time_seq}, time evolves from left to right and from top to bottom. We can see that three halos evolve under time and merge into a single, larger halo. This implies that our model is proficient in recovering halo evolution in in high resolution simulation generated based on LR simulations, which could save substantial computer resources. We also generate a static plot and overlap the trajectories of the halos onto the density plot. The top four massive halos within that $5 \hmpc$ box are selected and their trajectories are overlapped with the 3D density plot. The final position is denoted by an 'x' in the figure, and the redshift range is labeled at the lower left corner. In Figure \ref{fig:halo_traj} we can see that even though we do not have a hard constraint on each adjacent snapshot (such as applying a loss function in the time domain), the halos are moving smoothly. This suggests that the behavior of halos with fewer than 500 particles can also be learned from corresponding HR simulations.

\subsection{Full field statistics}
\label{subsection:full-field}

\subsubsection{Matter power spectrum}

\begin{figure}
    \centering
    \includegraphics[width=1.0\columnwidth]{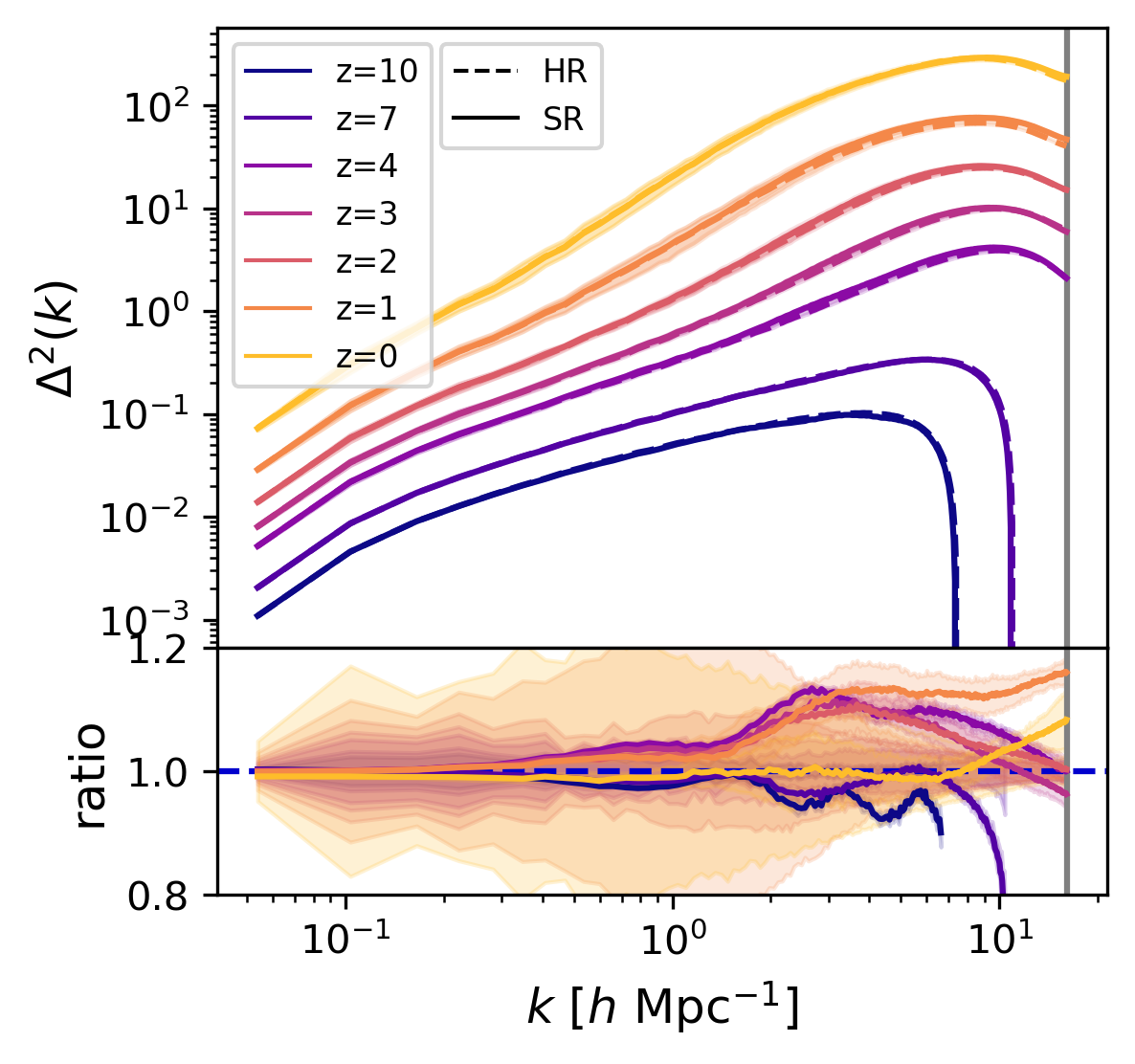}
    \caption{The matter power spectrum at integer redshifts of HR (dashed lines) and SR (solid lines). The upper panel shows matter power spectra while lower panel presents the ratio of SR and HR power spectra. Nyquist wavenumber is denoted by vertical solid gray line. The shaded area shows the 1$\sigma$ standard deviation measured from the 10 test sets. }
    \label{fig:matter_pwr}
\end{figure}

The matter power spectrum is a widely employed tool that characterises the amplitude of matter density fluctuations as a function of scale. It is the most common summary statistic used for the density field. It is defined as follows:
\begin{align*}
    P(k_i) = \frac{1}{N_i} \sum_{k_i < k < k_{i+1}} \frac{|\delta(k)|^2}{V}
\end{align*}
Where V is the volume of the simulation box and $\delta(k)$ is the Fourier transform of the overdensity field, which is defined as 
\begin{align*}
    \delta(x) = \frac{\rho(x) - \Bar{\rho}}{\Bar{\rho}}
\end{align*}
The $P(k)$ 
can be used as a metric to evaluate the accuracy of the SR field. In our work, we compare the dimensionless power spectrum $\Delta^2(k) \equiv \frac{k^3P(k)}{(2\pi)^3}$ of SR and HR fields at various redshifts.  The upper panel of Figure \ref{fig:matter_pwr} shows the matter power spectrum of HR (dashed lines) and SR (solid lines) at different redshifts, and the lower panel shows the ratio of the SR matter power spectrum to the HR matter power spectrum. The vertical dashed line  represents the Nyquist wavenumber, $k_{Nyq} = \frac{\pi N_{mesh}}{L_{box}}$, the scale below which the discrete Fourier transform cannot accurately capture information.

We compute the matter power spectrum for all the HR and SR simulations in our test set at integer redshifts. It is important to note that our training set does not contain integer redshifts; therefore, we utilize integer redshifts to assess the model's extrapolation capabilities in handling unseen data. The matter power spectra are generated using two distinct models: the first model is fine-tuned within the redshift range of z = 5 to z = 0, while the second model is fine-tuned for the redshift interval of z = 15 to z = 5. In figure \ref{fig:matter_pwr},  we can see that the SR power spectra successfully matches the HR results for small $k$ (large scales) within 5 percent. The power spectrum matches the power spectrum at large $k$ (small scales) within 20 percent at Nyquist wavenumber. The neural network successfully learned small scale structure directly from our target, the HR simulation. 
At high redshift $z = 7$ and $z = 10$, the matter power spectrum diminishes toward 0 rapidly, 
resulting in extreme values in the ratio plot at large k.

        
\subsection{Halo catalog analysis}
\label{subsection: halo-catalog}

\subsubsection{Halo Mass Function}
\begin{figure}
    \centering
    \includegraphics[width=1.0\columnwidth]{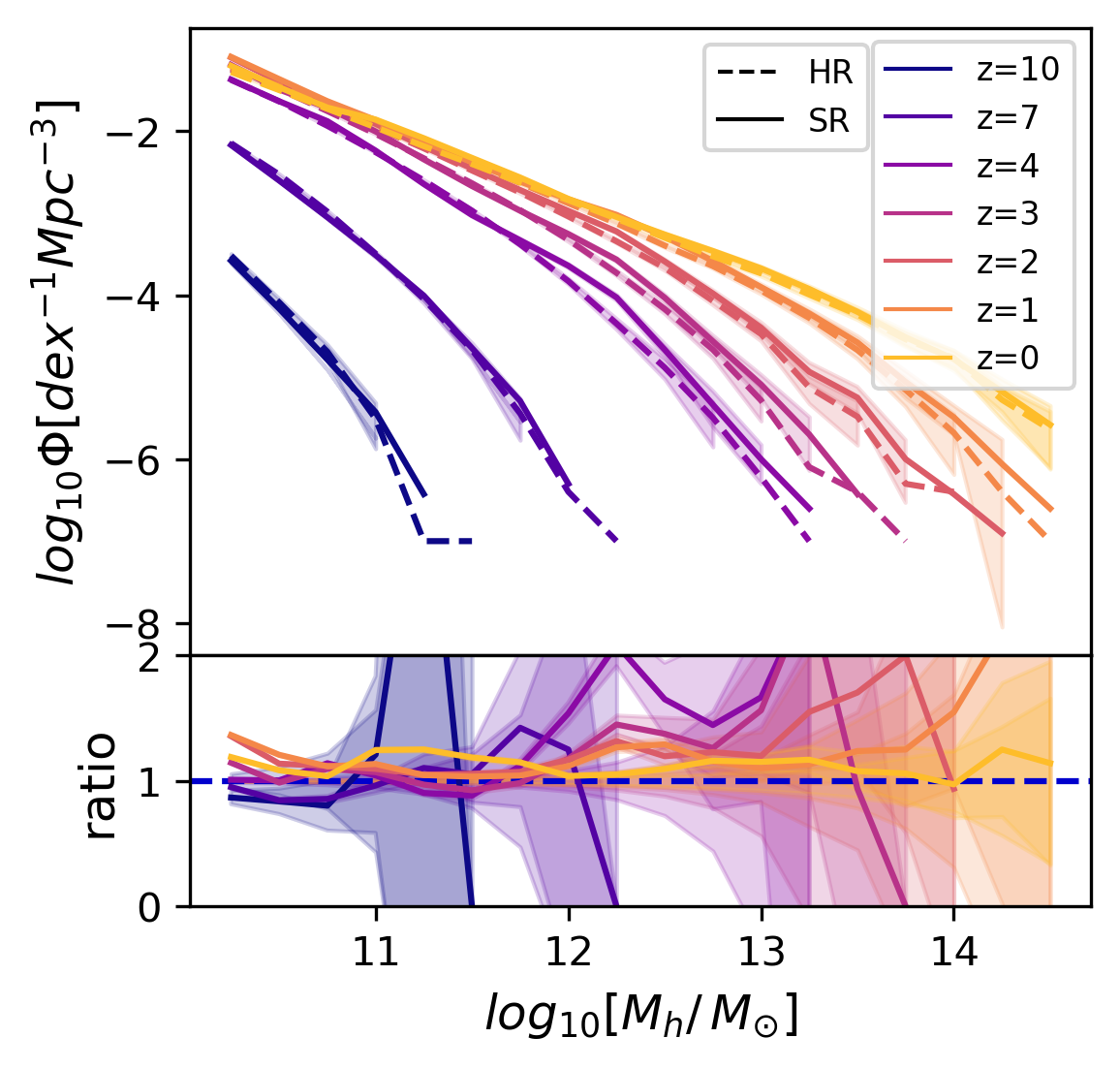}
    \caption{The halo mass function at different redshifts. Upper panel shows the halo mass function from SR  (solid lines) and HR (dashed lines) at various redshifts, and the lower panel is the ratio of the SR halo mass function to the HR equivalent. The shaded area shows the 1$\sigma$ standard deviation measured from the 10 test sets.}
    \label{fig:hmf}
\end{figure}

During the formation of cosmic structure,  some dark matter particles  become bound by the gravitational potential of halos into high-density structures. We use the friends-of-friends algorithm with linking length $b = 0.2$, meaning that particles within a faction of 0.2 of the simulation mean inter-particle separation are linked together.  Groups with a minimum of 32 particles are considered halos. The mass function of halos is defined as:
\begin{align}
    \phi = dn/d\log_{10}{M_h}
\end{align}
Where $dn$ is the comoving number density of halos within an infinitesimal logarithmic mass bin $d\log_{10}{M_h}$. This halo mass function is plotted in Figure \ref{fig:hmf}.

The neural network we have developed provides highly accurate predictions for the halo mass function at various redshifts, with a particular focus on the low-mass end at low redshift. The network achieves this precision by learning small-scale structures from the target high-resolution simulations instead of solely relying on the low-resolution input data. Our model successfully captures the small-scale structures and the halos with masses as low as $10^{11} \hmsun$. For high redshift($z = 10, 7$), the ratio plot shows large error bar at $10^{11} \hmsun$ and $10^{12} \hmsun$, which because only one or two halos are formed with large masss at high redshifts, and our SR model successfully predicts those halos. The halo mass function shows 50 percent discrepancy at $z = 4$ in region $10^{12} \hmsun to 10^{13} \hmsun$, this outcome is influenced by our training set sampling strategy, which is uniformly distributed in log(a). However, our result at z=0 takes advantage of this sampling strategy, successfully aligning with the HR halo mass function within 10 percent discrepancy.

\subsection{Merger trees}
The Rockstar code \citep{ROCKSTARpaper} is utilized to generate the merger tree history for the HR and generated SR fields. The algorithm employs the adaptive hierarchical refinement of FOF method to whole 6D phase space, which are then progressively refined to determine the subhalos. The particles are assigned to the closest halo, and unbound particles are removed. The halo properties are then calculated, with the halo containing the maximum number of particles from the previous time step considered its descendant.

\begin{figure*}
\centering
  \includegraphics[width=0.33\textwidth]{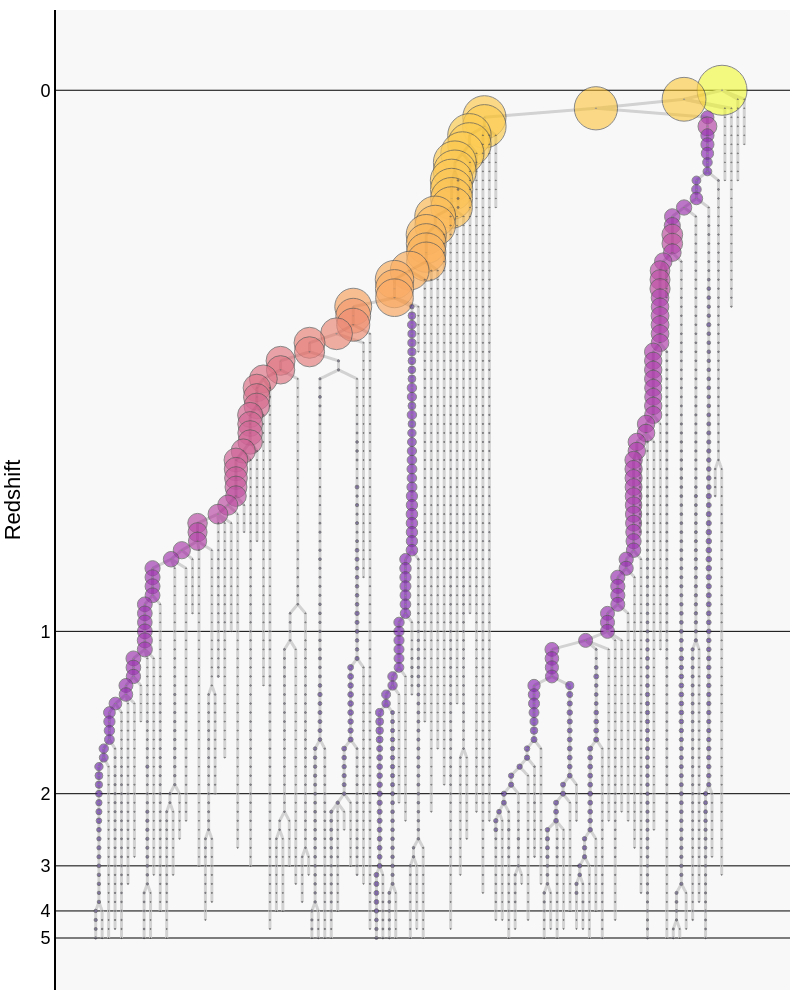}
  \includegraphics[width=0.33\textwidth]{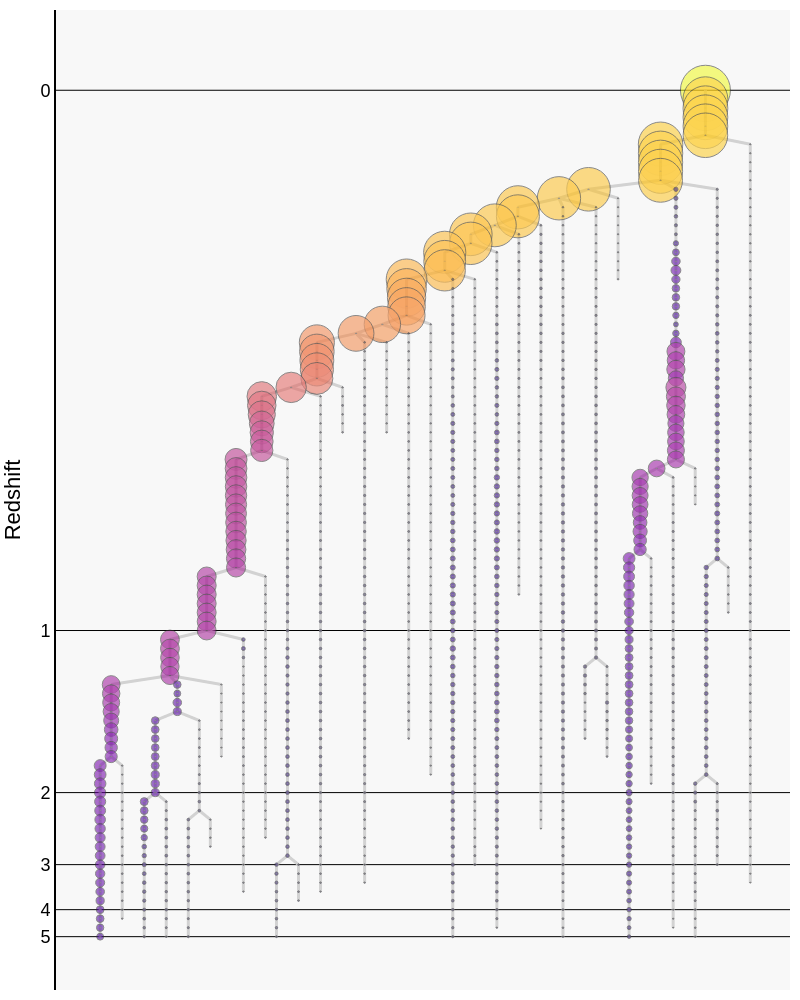}
  \includegraphics[width=0.33\textwidth]{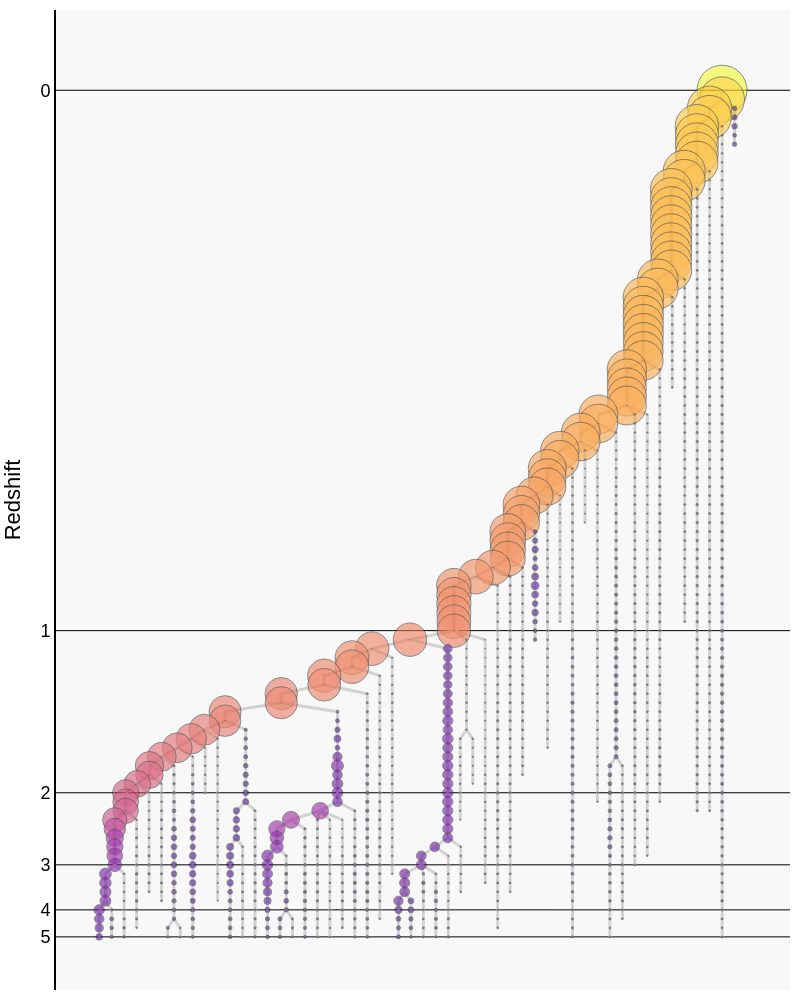}\\
  \includegraphics[width=0.33\textwidth]{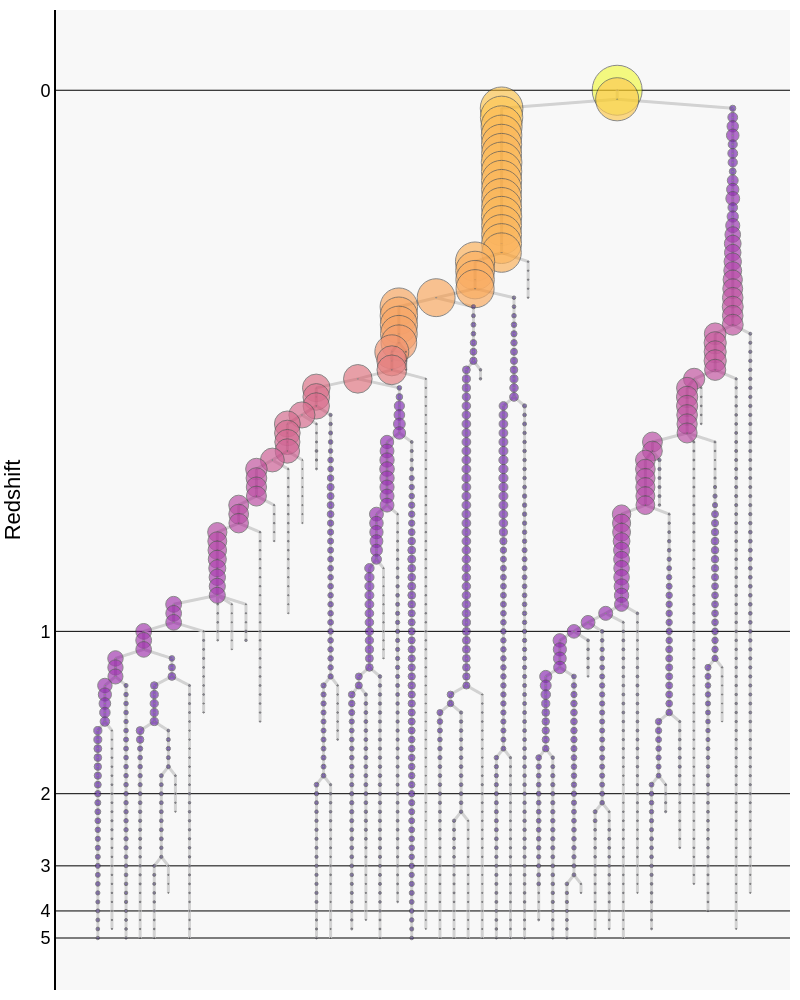}
  \includegraphics[width=0.33\textwidth]{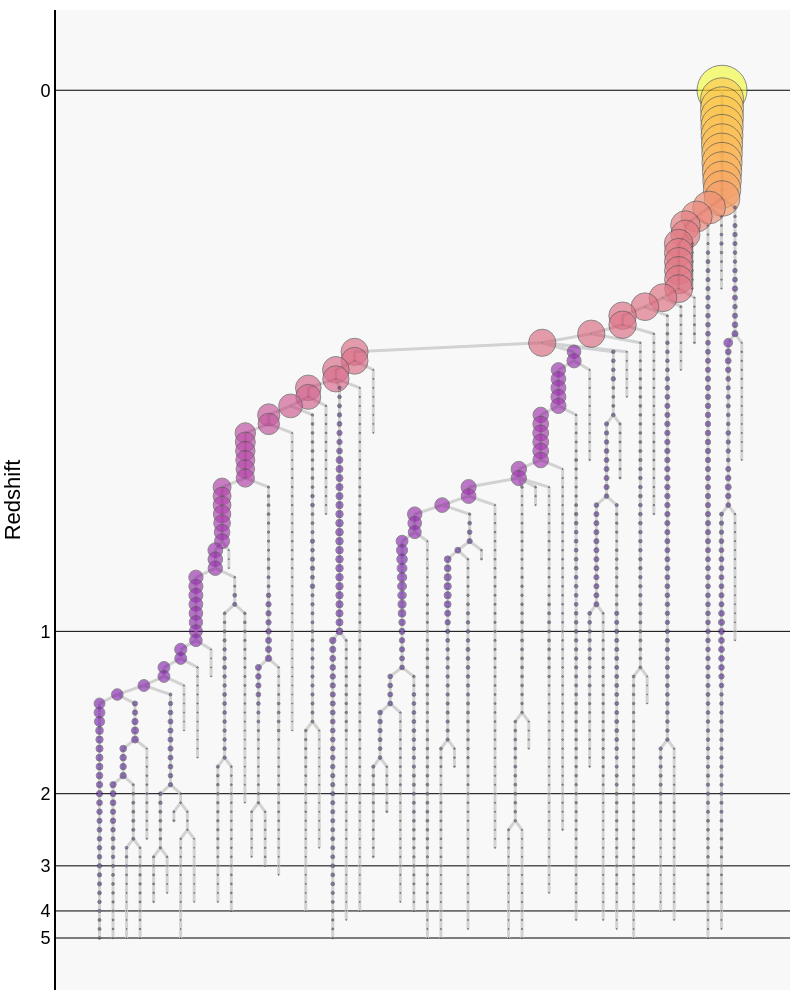}
  \includegraphics[width=0.33\textwidth]{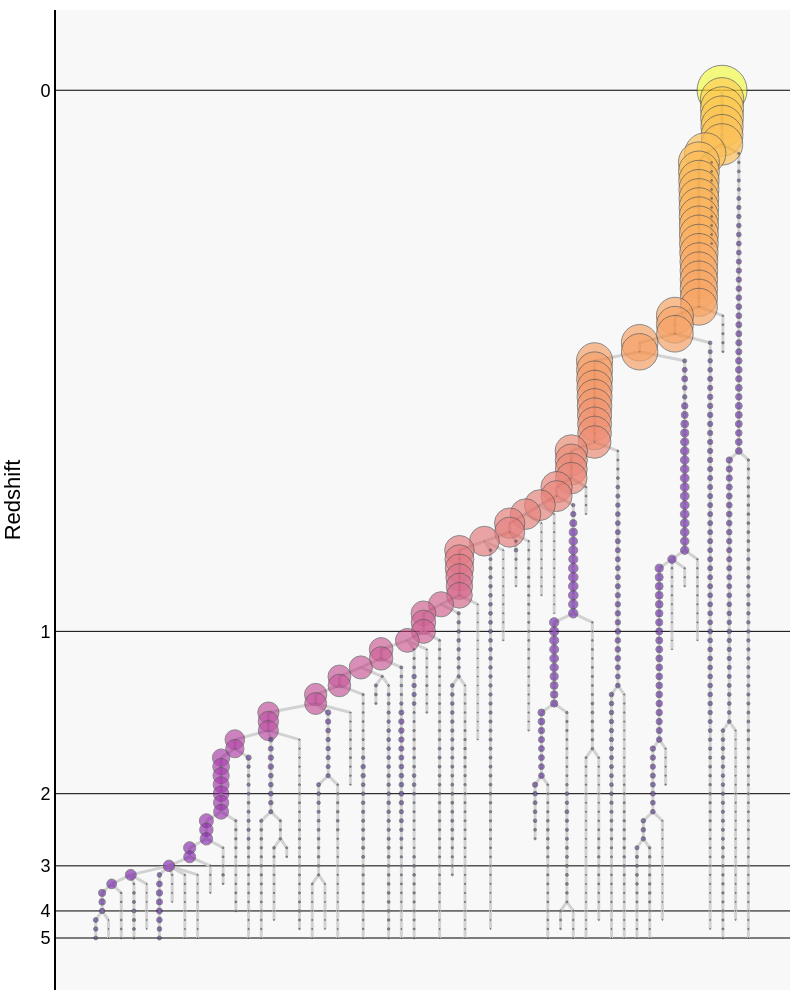}
  \caption{Examples of generated merger trees, where the upper 3 merger trees are from the HR field and the lower 3 merger trees are from the SR field. Time is depicted as progressing from bottom to top, covering the redshift range of $z = 5$ to $z = 0$. The radius and color of each dot is proportional to the halo mass.} 
  \label{fig:merger-tree}
\end{figure*}

In our study, we use 95 time-consistent snapshots from LR to create SR and employ the consistent-trees algorithm to create the merger trees. Merger trees are constructed by using particle IDs to match halos between snapshots and parent-child relationships are determined. We generate 3 different time evolution test sets from LR simulations and compare them with corresponding HR simulations.

Figure \ref{fig:merger-tree} shows examples of massive halo merger trees, with the upper panels taken from the HR field and the lower from the SR field, with the bottom to top progression representing redshift $z=5$ to redshift $z=0$. The size of the circles is proportional to the square root of the halo mass. As time evolves, the halo gains mass while smaller halos merge into the main progenitor (the most massive halo at each snapshot). We can find that HR trees shows mass fluctuations, and our merger trees derived from SR field successfully capture these dynamics.To further compare merger trees from HR and SR, we utilize additional statistical methods to compare SR and HR merger trees.

\subsubsection{Main branch length}
\label{mbl}
\begin{figure}
    \centering
    \includegraphics[width=1.0\columnwidth]{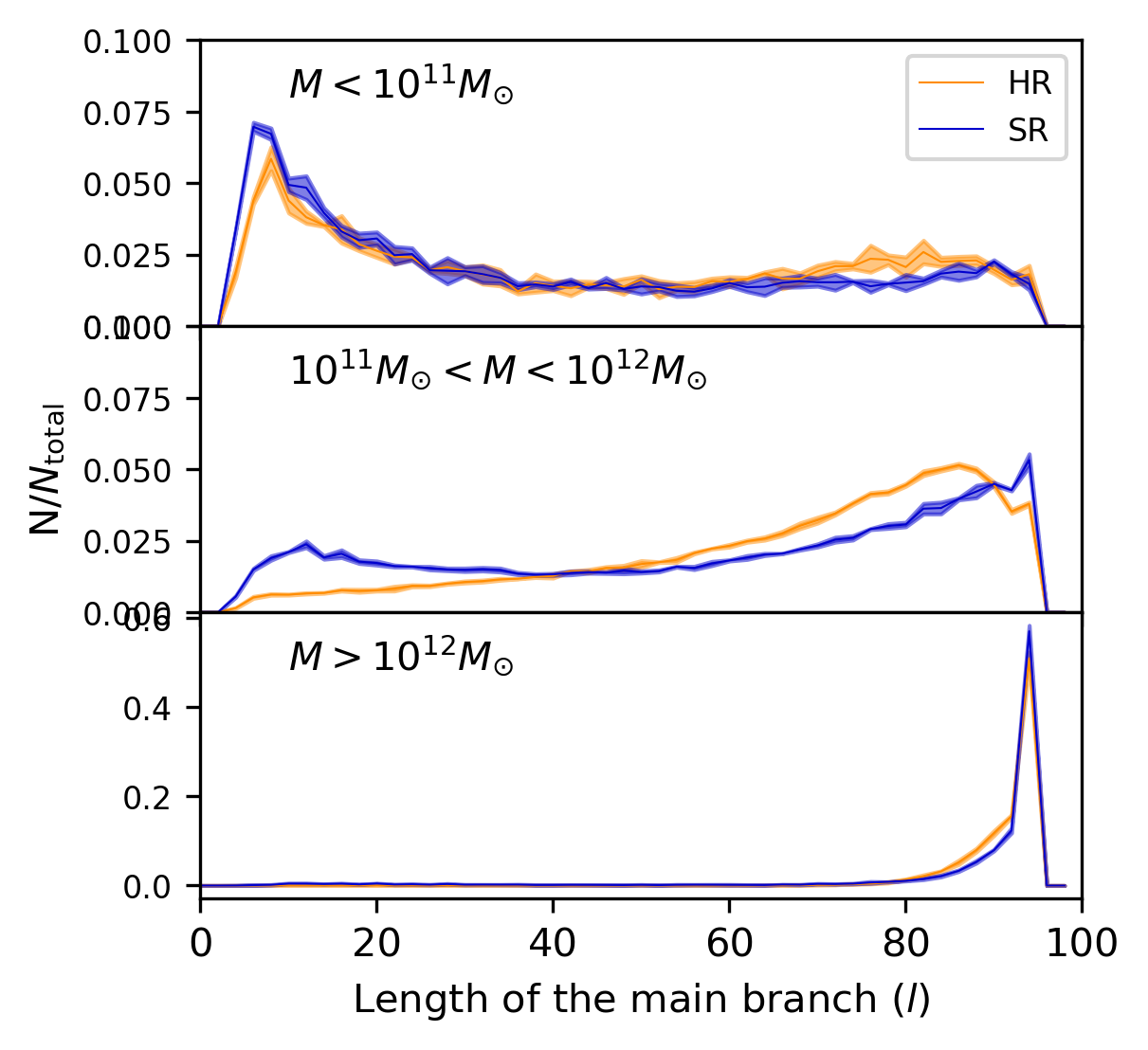}
    \caption{
    The distribution of the main branch length, $l$, of merger trees at redshift $z=0$.  The fraction of merger trees in each $l$-bin is displayed on the y-axis, and the x-axis is the length of the main branch in number of snapshots. The halos are divided into three mass bins, as shown in the upper, middle and lower panels. The shaded area shows the 1$\sigma$ standard deviation measured from the 3 test sets.
    }
    \label{fig:mbl}
\end{figure}

One of the most straightforward attributes of a tree is the length of the main branch. This measures the extent to which a halo can be traced back in time, initiating, in our case, at $z = 0$. The length of the main branch for a particular halo is determined by the number of snapshots where the most massive progenitor of that halo exists \citep{sussing}.
Figure \ref{fig:mbl} shows the distribution of main branch length for halos in different mass ranges. The halo mass ranges are divided into three regions, with the upper panel showing halos with mass less than $10^{11} \hmsun$, the middle panel displaying halos with mass ranging from $10^{11} \hmsun$ to $10^{12} \hmsun$, and the lower panel showing halos with mass greater than $10^{12} \hmsun$. The solid blue lines represent the main branch length distribution calculated from the SR field. The dashed black lines are the distribution derived from HR simulations, with the same initial conditions for the LR which serves as the input for the SR field.
In general, larger halos tend to have a longer merger history compared to smaller halos, as they are typically formed through the merging of multiple smaller halos over time. 
For halos with mass $M < 10^{11} \hmsun$, we observe that the both SR and HR have many short branch halos, meaning that many of them are born roughly 10 snapshots ago. Our SR model succesfully capture that with consistency at mass region $M < 10^{11} \hmsun$. Similarly in mass region $M > 10^{12} \hmsun$, the massive halos tend to have longer history. Remarkably, the SR and HR main branch length values exhibit essentially good agreement for halos of $M < 10^{11}\hmsun$. For larger mass halos within range  $10^{11} < M < 10^{12} \hmsun$, the SR field predicts a higher number of halos with a shorter merger histories compared to HR. 
This discrepancy may arise from unresolved small-scale structures and associated randomness in particles, leading to incorrect resolution of halo histories and thereby generating more halos with short merger histories. Additionally, it could be that the SR model is generating more noisy small-scale structures, causing the halo to appear later than expected. Additionally, in Figure 3 (middle panel) of the paper cited as \citep{sussing}, a discrepancy between different tree building algorithms is comparable to the difference we have between HR and SR model in our study. This suggests that while our SR model does introduce additional source of error, its performance is comparable to variation in different tree building algorothms.

\subsubsection{Number of direct ancestors}

\begin{figure}
    \centering
    \includegraphics[width=1.0\columnwidth]{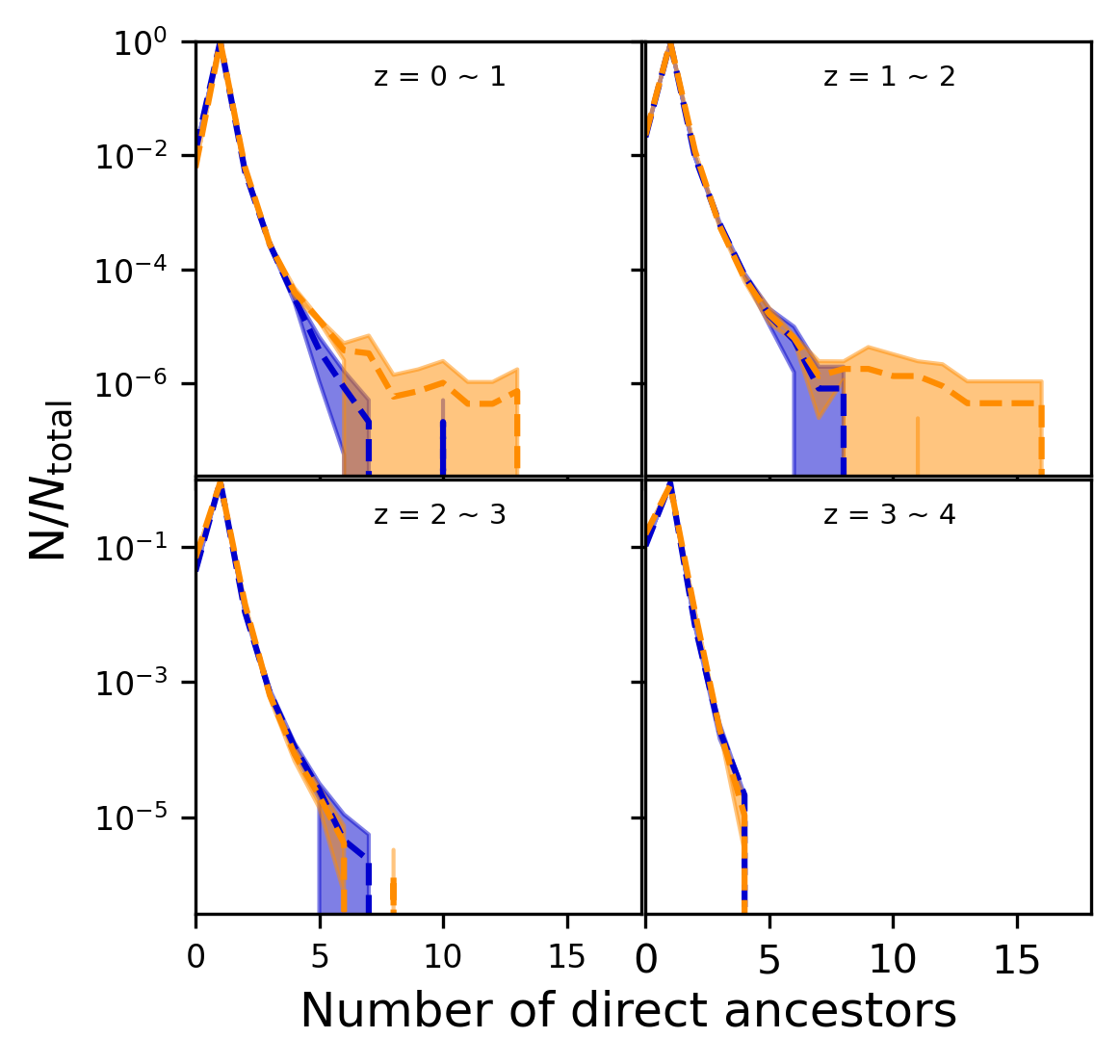}
    \caption{Plots of the number of direct ancestors, using halos from different redshift ranges. The blue curve are counts of halos (tree nodes) calculated from SR fields and the orange histograms are from HR simulations with the same initial conditions and random seed. The shaded area shows the 1$\sigma$ standard deviation measured from the 3 test sets.}
    \label{fig:direct_prog}
\end{figure}

The number of direct ancestors at each node in the merger tree is also a subject of interest, as it provides a direct measurement of the width distribution of all the merger trees \citep{sussing} (cf. the length of trees, quantified in Section \ref{mbl}). In Figure \ref{fig:direct_prog} we plot the number of tree nodes with direct ancestors, for both HR and SR, including all halos taken from each of four different redshift ranges. 
The maximum number of direct ancestors for halos in different redshift regions was investigated in both HR and SR simulations. While the precise value of this statistic depends on the spacing between snapshots, it is still interesting to compare HR and SR results since the SR fields were generated from the corresponding LR simulations. Halos were classified based on the number of direct ancestors, with most having only one direct ancestor in both HR and SR. As redshift increases, halos tend to have fewer direct ancestors. Our SR model successfully predicts a statistically accurate number of merger events across different redshift ranges.

\subsection{Mass growth along the halo main branch}
\begin{figure}
    \centering
    \includegraphics[width=1.0\columnwidth]{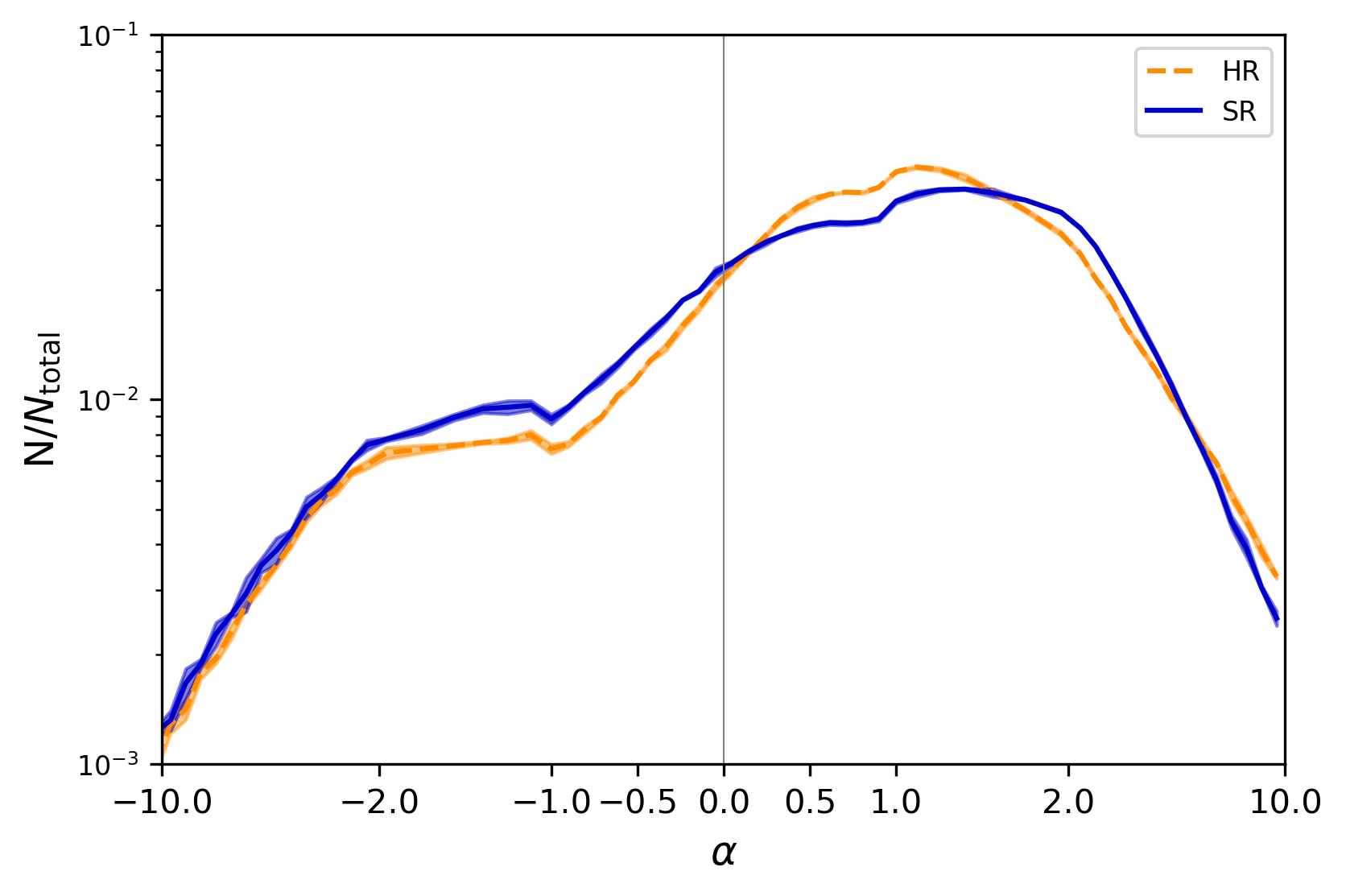}
    \caption{Distribution function of logarithmic mass($M_{200c}$) growth along halo main branches, for all pairs of halos where both masses exceed $10^{12} \hmsun$}
    \label{fig:mass_growth}
\end{figure}

The logarithmic growth rate of main branch halos, denoted as $dlog M/dlogt$, can be approximated by the following expression (\citealt{sussing}),
\begin{align}
    \frac{d log M}{d log t} \approx \alpha_M (A, B) = \frac{(t_{k+1} + t_{k})(M_{k+1} - M_{k})}{(t_{k+1} - t_{k})(M_{k+1} + M_{k})}
    \label{alpha}
\end{align}
Here, $M_k$ and $M_{k+1}$ are the masses of a halo and its descendent at
times $t_k$ and $t_{k+1}$, respectively. The distribution function of $\alpha_M$ is
shown in Figure \ref{fig:mass_growth} for every pair of main-branch halos for which
the mass of each exceeds $10^{12} \hmsun$. A positive $\alpha$ value implies mass growth from $t_k$ to $t_{k+1}$ and negative $\alpha$ value implies mass loss from $t_k$ to $t_{k+1}$. 

As seen in Figure \ref{fig:mass_growth}, most of the time halos are gaining mass, but there are also significant periods of mass loss. However, significant mass loss is unphysical and is due to misidentifications occurring during the halo identification and tree construction phases \cite[see][for more discussion]{sussing}. Most importantly, we can see that the distribution of halo mass growth for HR and SR simulations shows consistency. The SR curve display a higher peak of halos losing mass with $\alpha $ from -2 to -1 and a shifted peak from $\alpha = 1$ to $\alpha$ close to 2. This is impacted by the presence of noisy small-scale structures and the collection of particles generated by SR model surrounding the corresponding LR particle chunk, which can lead the halo finder and tree builder to predict mass loss more frequently, or, predict that halos are accruing more mass than they should. This is limited by the super-resolution model which can not have visually sharp structure and good generalization at the same time.

\subsection{Mass fluctuations of halo main branches}

We would like to quantify the mass fluctuations that come from the natural growth process or from halo misidentification and use this statitic to compare our SR and HR simulations. Again following \citep{sussing}, we use the statistic $\xi({k})$ for this purpose:

\begin{align*}
    \xi(k) = \texttt{arctan} \alpha_M(k, k+1) - \texttt{arctan} \alpha_M(k-1, k)
\end{align*}

Here, $\alpha_M$ is defined as in Eq.~\ref{alpha},  and $k-1, k, k+1$ represent consecutive timesteps. 

\begin{figure}
    \centering
    \includegraphics[width=1.0\columnwidth]{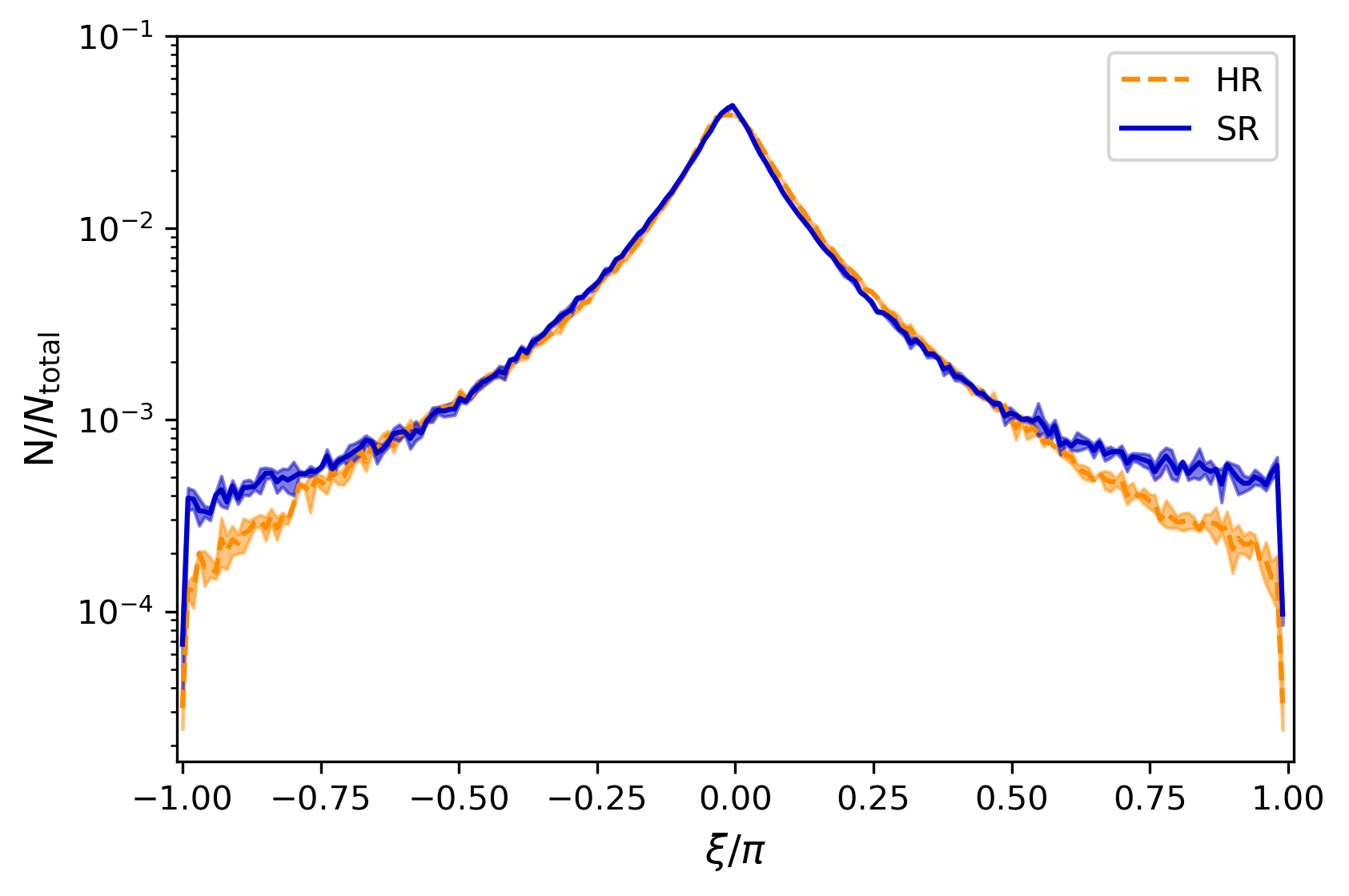}
    \caption{Distribution function of logarithmic mass growth along halo main branches, all pairs of halos for both mass exceed $10^{12} \hmsun$}
    \label{fig:fluctuations}
\end{figure}

Physically, $\xi(k)$ measures the change in the slope of the mass accretion rate between two consecutive steps and thus ranges
from $\pi$ to $-\pi$. Here the extreme cases are confined to the region where $\xi/\pi$ is close to  -1 and 1, but at these extremes the statistic indicates that the halo is losing and then gaining (or gaining then losing) mass over two consecutive time steps. 

In Figure \ref{fig:fluctuations}, the mass fluctuations measured from SR and HR models matches well around $\xi = 0$, and so indicates that the SR model successfully captures the process of natural mass gain and loss during halo evolution. The discrepancies that are seen represent a very small fraction of halos: the SR curve at -1 and 1 is higher than the HR curve, meaning more extreme cases than HR field. This is also consistent with the conclusion from Figure \ref{fig:mass_growth}.

\section{Discussion}
\label{section4:Discussion}
The application of super-resolution (SR) techniques to the 6D phase space of cosmological simulations has been explored in previous works \citep{AI1,AI2}, where neural networks (NNs) were trained on a single redshift, limiting the generated SR field to only that redshift. In contrast, our SR model, once trained is able to generate a full 6D super-resolution field at any redshift required, using the scale factor as a style parameter to control the output Lagrangian field (displacement and velocity). Notably, by using data from an LR simulation consisting of a set of snapshots ordered in time, we can generate the time evolution of SR field, where the large-scale structure is naturally consistent with the LR input and small-scale structures also have time consistency but can be controlled via random seeds. Our generated SR field can be treated in exactly the same fashion as the output of an HR simulation with distinguishable tracer particles, allowing for the construction of merger histories through halo finders and tree-building codes.

In this study, we implemented our SR model to produce outputs with 512 times higher mass resolution than LR simulations at various redshifts. To validate the accuracy of our SR field, we compared the matter power spectrum of the density field with HR simulations and found that our model can match the power spectra within 5 percent on small $k$. For large $k$, the maximum deviation observed lies within a margin of 20 percent. This result is a significant improvement compared to LR simulations that lack small-scale structures. To evaluate the non-Gaussianity of the SR field and the time evolution of dark matter particles, we utilized the FOF algorithm to calculate the halo mass function and ROCKSTAR+Consistent-Trees to obtain the merger trees.

The halo mass function was computed for both the SR and HR fields, and the results show good agreement (within 20 percent) for all halos with masses down to $10^{11} \hmsun$ in the SR field. For $z = 4$ the difference is around $50$ percent. However, by $z = 0$, the discrepancy is $< 10$ percent.

The merger trees also reveal a high level of similarity between the SR and HR simulations. This demonstrates the effectiveness of our SR approach for capturing the time evolution of a dark matter simulation. Furthermore, we also investigate different statistics to measure the differences between SR and HR results. These measurement indicate that the SR model can create statistical accurate mock halo catalogs which can provide insight into the formation and evolution of structures in the universe. Overall, this work has demonstrated the potential of using neural networks for generating the full phase space from LR simulations and should have wide implications for studying structure formation in cosmology.

In this study, we have expanded on the findings of our previous work (paper II) to encompass a wider range of redshifts. Our current efforts are focused on further refining the super-resolution technique by imposing stricter constraints on the time domain. Additionally, we plan to incorporate the use of additional style parameters that represent cosmological parameters, or perhaps simulation parameters (e.g., resolution enhancement factor) in order to enable wider applications of the SR technique in cosmology.

Our study demonstrates that the accuracy of the SR model strongly depends on the selection of particular statistical metrics, which correlate with different applications. Consequently, additional constraints can be implemented and incorporated into the neural network. Alternatively, we could consider adopting a different model in the future.

\section{Summary and Conclusion}
\label{section5:Conclusion}
In this work, we present a super-resolution model that can generate the full 6D phase space displacement and velocity fields represented by tracer particles, equivalent to the output from a real cosmological simulation. Our model is conditioned on low-resolution simulations and takes as an input style parameter the required cosmic scale factor. By doing so, our model enhances the simulation resolution by generating 512 times more tracer particles at various redshifts. 
We compare power spectra and halo mass functions obtained from our novel super-resolution (SR) models to those from the corresponding cosmological $N$-body simulations. 
We first validate the effectiveness of our super-resolution (SR) model by examining statistics of the entire dark matter field. We demonstrate that our generated SR fields match the power spectra of true high-resolution (HR) simulations to within a 20 percentage level.

In addition to validating the statistical properties of our SR model, we demonstrate its ability to generate visually authentic small-scale structures that are well beyond the resolution of the LR input.  We also present a visualization of the temporal evolution of the SR field.
We conduct an analysis of dark matter halos within the generated SR fields. Our results demonstrate that our SR model can produce small-scale structures that possess a visually realistic appearance, surpassing the resolution of the LR input and exhibiting statistical agreement with those derived from direct $N$-body simulations. The quantitative analysis of the abundance of the halo population shows that the SR field exhibits good agreement with the HR field, with differences of less than 20 percent at most redshifts. At $z = 4$, we observe an overshoot in the halo mass function of approximately 40 percent around $10^{12} \hmsun$ to $10^{13} \hmsun$.

To investigate the time evolution of the SR field, we generate 95 snapshots of the SR field from a time-consistent LR simulation. This enables us to create realistic merger trees for halos at redshift $z = 0$, as well as track the evolution of the final halo. Due to differences in small-scale structures between the SR and HR fields, direct comparison of individual merger trees is not meaningful. Therefore, we employ the main branch length and branching ratio to statistically compare the population of generated merger trees from the SR and HR fields. We evaluate the mass fluctuation along the main branch and our SR model predicts higher percentage of halos losing mass and a shifted peak indicates that halos are gaining more mass.
Our findings demonstrate that the SR model is capable of generating merger histories that are solely dependent on the time-consistent LR input. This indicates the potential for studying time evolution of cosmic structure in detail using only LR simulations as an input. One important application of this work is the creation of mock catalogs from large scale surveys. The generation of realistic halo merger trees plays an important role in studying galaxy formation. Building semi-analytic models based on accurate halo identification and well-construct merger trees is quite costly through conventional high-resolution large boxsize N-body simulations. However, with the use of SR model, we are able to construct these realistic merger trees in a more time and resource efficient manner. 

\section*{Data Availability}
Our framework for training the super-resolution (SR) model is available at \url{ https://github.com/sagasv5-xw/map2map} on the 'styled srsgan' branch. This PyTorch-based framework is a general-purpose tool for transforming field data. The trained SR models and the pipeline for generating SR fields can also be found in the same branch. The training and test data sets generated during this work will be made available upon reasonable request to the corresponding author.

\section*{Acknowledgements}
This research is part of the Frontera computing project at the Texas Advanced Computing Center. Frontera is made possible by NSF award OAC-1818253.
TDM acknowledges funding from NSF ACI-1614853, NSF AST-1616168, NASA ATP 19-ATP19-0084 and 80NSSC20K0519.
TDM and RACC also acknowledge funding from NASA ATP 80NSSC18K101, and NASA ATP NNX17AK56G, and RACC, NSF AST-1909193.
SB was supported by NASA ATP 80NSSC22K1897.
This work was also supported by the NSF AI Institute: Physics of the Future, NSF PHY-2020295.
The Flatiron Institute is supported by the Simons Foundation.
We also acknowledge the code packages used in this work:
The simulations for training and testing is run with \texttt{MP-Gadget} (\url{https://github.com/MP-Gadget/MP-Gadget}).
Visualization in this work is performed with open source code \texttt{gaepsi2} (\url{https://github.com/rainwoodman/gaepsi2}) and package \texttt{pltoly} \citep{plotly}.
Data and catalog analysis in this work is performed with open-source
software \texttt{PyTorch}\citep{pytorch}, \texttt{nbodykit}\citep{Hand2018}, \texttt{ytree}\citep{ytree}, and \texttt{ROCKSTAR} halo finder \citep{ROCKSTARpaper}



\bibliographystyle{mnras}

\bibliography{bib.bib}

\begin{thebibliography}{}
\makeatletter
\relax
\def\mn@urlcharsother{\let\do\@makeother \do\$\do\&\do\#\do\^\do\_\do\%\do\~}
\def\mn@doi{\begingroup\mn@urlcharsother \@ifnextchar [ {\mn@doi@}
  {\mn@doi@[]}}
\def\mn@doi@[#1]#2{\def\@tempa{#1}\ifx\@tempa\@empty \href
  {http://dx.doi.org/#2} {doi:#2}\else \href {http://dx.doi.org/#2} {#1}\fi
  \endgroup}
\def\mn@eprint#1#2{\mn@eprint@#1:#2::\@nil}
\def\mn@eprint@arXiv#1{\href {http://arxiv.org/abs/#1} {{\tt arXiv:#1}}}
\def\mn@eprint@dblp#1{\href {http://dblp.uni-trier.de/rec/bibtex/#1.xml}
  {dblp:#1}}
\def\mn@eprint@#1:#2:#3:#4\@nil{\def\@tempa {#1}\def\@tempb {#2}\def\@tempc
  {#3}\ifx \@tempc \@empty \let \@tempc \@tempb \let \@tempb \@tempa \fi \ifx
  \@tempb \@empty \def\@tempb {arXiv}\fi \@ifundefined
  {mn@eprint@\@tempb}{\@tempb:\@tempc}{\expandafter \expandafter \csname
  mn@eprint@\@tempb\endcsname \expandafter{\@tempc}}}

\bibitem[\protect\citeauthoryear{{Arjovsky}, {Chintala}  \&
  {Bottou}}{{Arjovsky} et~al.}{2017}]{wgan}
{Arjovsky} M.,  {Chintala} S.,   {Bottou} L.,  2017, \mn@doi [arXiv e-prints]
  {10.48550/arXiv.1701.07875}, \href
  {https://ui.adsabs.harvard.edu/abs/2017arXiv170107875A} {p. arXiv:1701.07875}

\bibitem[\protect\citeauthoryear{{Bagla}}{{Bagla}}{2002}]{bagla02}
{Bagla} J.~S.,  2002, \mn@doi [Journal of Astrophysics and Astronomy]
  {10.1007/BF02702282}, \href
  {https://ui.adsabs.harvard.edu/abs/2002JApA...23..185B} {23, 185}

\bibitem[\protect\citeauthoryear{{Behroozi}, {Wechsler}  \& {Wu}}{{Behroozi}
  et~al.}{2013}]{ROCKSTARpaper}
{Behroozi} P.~S.,  {Wechsler} R.~H.,   {Wu} H.-Y.,  2013, \mn@doi [\apj]
  {10.1088/0004-637X/762/2/109}, \href
  {https://ui.adsabs.harvard.edu/abs/2013ApJ...762..109B} {762, 109}

\bibitem[\protect\citeauthoryear{{Berger} \& {Stein}}{{Berger} \&
  {Stein}}{2019}]{berger2019}
{Berger} P.,  {Stein} G.,  2019, \mn@doi [\mnras] {10.1093/mnras/sty2949},
  \href {https://ui.adsabs.harvard.edu/abs/2019MNRAS.482.2861B} {482, 2861}

\bibitem[\protect\citeauthoryear{{Bernardini}, {Mayer}, {Reed}  \&
  {Feldmann}}{{Bernardini} et~al.}{2020}]{Bernardini2020}
{Bernardini} M.,  {Mayer} L.,  {Reed} D.,   {Feldmann} R.,  2020, \mn@doi
  [\mnras] {10.1093/mnras/staa1911}, \href
  {https://ui.adsabs.harvard.edu/abs/2020MNRAS.496.5116B} {496, 5116}

\bibitem[\protect\citeauthoryear{{Dai} \& {Seljak}}{{Dai} \&
  {Seljak}}{2021}]{Dai2021}
{Dai} B.,  {Seljak} U.,  2021, \mn@doi [Proceedings of the National Academy of
  Science] {10.1073/pnas.2020324118}, \href
  {https://ui.adsabs.harvard.edu/abs/2021PNAS..11820324D} {118, 2020324118}

\bibitem[\protect\citeauthoryear{{Dvorkin} et~al.,}{{Dvorkin}
  et~al.}{2022}]{MLincosmo}
{Dvorkin} C.,  et~al., 2022, \mn@doi [arXiv e-prints]
  {10.48550/arXiv.2203.08056}, \href
  {https://ui.adsabs.harvard.edu/abs/2022arXiv220308056D} {p. arXiv:2203.08056}

\bibitem[\protect\citeauthoryear{{Goodfellow}, {Pouget-Abadie}, {Mirza}, {Xu},
  {Warde-Farley}, {Ozair}, {Courville}  \& {Bengio}}{{Goodfellow}
  et~al.}{2014}]{goodfellow2014GAN}
{Goodfellow} I.~J.,  {Pouget-Abadie} J.,  {Mirza} M.,  {Xu} B.,  {Warde-Farley}
  D.,  {Ozair} S.,  {Courville} A.,   {Bengio} Y.,  2014, \mn@doi [arXiv
  e-prints] {10.48550/arXiv.1406.2661}, \href
  {https://ui.adsabs.harvard.edu/abs/2014arXiv1406.2661G} {p. arXiv:1406.2661}

\bibitem[\protect\citeauthoryear{{Gulrajani}, {Ahmed}, {Arjovsky}, {Dumoulin}
  \& {Courville}}{{Gulrajani} et~al.}{2017}]{wgan-gp}
{Gulrajani} I.,  {Ahmed} F.,  {Arjovsky} M.,  {Dumoulin} V.,   {Courville} A.,
  2017, \mn@doi [arXiv e-prints] {10.48550/arXiv.1704.00028}, \href
  {https://ui.adsabs.harvard.edu/abs/2017arXiv170400028G} {p. arXiv:1704.00028}

\bibitem[\protect\citeauthoryear{{Hand}, {Feng}, {Beutler}, {Li}, {Modi},
  {Seljak}  \& {Slepian}}{{Hand} et~al.}{2018}]{Hand2018}
{Hand} N.,  {Feng} Y.,  {Beutler} F.,  {Li} Y.,  {Modi} C.,  {Seljak} U.,
  {Slepian} Z.,  2018, \mn@doi [\aj] {10.3847/1538-3881/aadae0}, \href
  {https://ui.adsabs.harvard.edu/abs/2018AJ....156..160H} {156, 160}

\bibitem[\protect\citeauthoryear{{He}, {Zhang}, {Ren}  \& {Sun}}{{He}
  et~al.}{2015}]{resnet}
{He} K.,  {Zhang} X.,  {Ren} S.,   {Sun} J.,  2015, \mn@doi [arXiv e-prints]
  {10.48550/arXiv.1512.03385}, \href
  {https://ui.adsabs.harvard.edu/abs/2015arXiv151203385H} {p. arXiv:1512.03385}

\bibitem[\protect\citeauthoryear{{He}, {Li}, {Feng}, {Ho}, {Ravanbakhsh},
  {Chen}  \& {P{\'o}czos}}{{He} et~al.}{2019}]{He_2019}
{He} S.,  {Li} Y.,  {Feng} Y.,  {Ho} S.,  {Ravanbakhsh} S.,  {Chen} W.,
  {P{\'o}czos} B.,  2019, \mn@doi [Proceedings of the National Academy of
  Science] {10.1073/pnas.1821458116}, \href
  {https://ui.adsabs.harvard.edu/abs/2019PNAS..11613825H} {116, 13825}

\bibitem[\protect\citeauthoryear{{Hinshaw} et~al.,}{{Hinshaw}
  et~al.}{2013}]{hinshaw13}
{Hinshaw} G.,  et~al., 2013, \mn@doi [\apjs] {10.1088/0067-0049/208/2/19},
  \href {https://ui.adsabs.harvard.edu/abs/2013ApJS..208...19H} {208, 19}

\bibitem[\protect\citeauthoryear{{Hopkins}, {Kere{\v{s}}}, {O{\~n}orbe},
  {Faucher-Gigu{\`e}re}, {Quataert}, {Murray}  \& {Bullock}}{{Hopkins}
  et~al.}{2014}]{fire}
{Hopkins} P.~F.,  {Kere{\v{s}}} D.,  {O{\~n}orbe} J.,  {Faucher-Gigu{\`e}re}
  C.-A.,  {Quataert} E.,  {Murray} N.,   {Bullock} J.~S.,  2014, \mn@doi
  [\mnras] {10.1093/mnras/stu1738}, \href
  {https://ui.adsabs.harvard.edu/abs/2014MNRAS.445..581H} {445, 581}

\bibitem[\protect\citeauthoryear{Inc.}{Inc.}{2015}]{plotly}
Inc. P.~T.,  2015, Collaborative data science, \url {https://plot.ly}

\bibitem[\protect\citeauthoryear{{Isola}, {Zhu}, {Zhou}  \& {Efros}}{{Isola}
  et~al.}{2016}]{isola2017image}
{Isola} P.,  {Zhu} J.-Y.,  {Zhou} T.,   {Efros} A.~A.,  2016, \mn@doi [arXiv
  e-prints] {10.48550/arXiv.1611.07004}, \href
  {https://ui.adsabs.harvard.edu/abs/2016arXiv161107004I} {p. arXiv:1611.07004}

\bibitem[\protect\citeauthoryear{{Jamieson}, {Li}, {Alves de Oliveira},
  {Villaescusa-Navarro}, {Ho}  \& {Spergel}}{{Jamieson}
  et~al.}{2022}]{jamieson2022field}
{Jamieson} D.,  {Li} Y.,  {Alves de Oliveira} R.,  {Villaescusa-Navarro} F.,
  {Ho} S.,   {Spergel} D.~N.,  2022, \mn@doi [arXiv e-prints]
  {10.48550/arXiv.2206.04594}, \href
  {https://ui.adsabs.harvard.edu/abs/2022arXiv220604594J} {p. arXiv:2206.04594}

\bibitem[\protect\citeauthoryear{{Karras}, {Laine}, {Aittala}, {Hellsten},
  {Lehtinen}  \& {Aila}}{{Karras} et~al.}{2019}]{stylegan2}
{Karras} T.,  {Laine} S.,  {Aittala} M.,  {Hellsten} J.,  {Lehtinen} J.,
  {Aila} T.,  2019, \mn@doi [arXiv e-prints] {10.48550/arXiv.1912.04958}, \href
  {https://ui.adsabs.harvard.edu/abs/2019arXiv191204958K} {p. arXiv:1912.04958}

\bibitem[\protect\citeauthoryear{{Kodi Ramanah}, {Charnock},
  {Villaescusa-Navarro}  \& {Wandelt}}{{Kodi Ramanah}
  et~al.}{2020}]{KodiRamanah2020}
{Kodi Ramanah} D.,  {Charnock} T.,  {Villaescusa-Navarro} F.,   {Wandelt}
  B.~D.,  2020, \mn@doi [\mnras] {10.1093/mnras/staa1428}, \href
  {https://ui.adsabs.harvard.edu/abs/2020MNRAS.495.4227K} {495, 4227}

\bibitem[\protect\citeauthoryear{{Laureijs} et~al.,}{{Laureijs}
  et~al.}{2011}]{euclid}
{Laureijs} R.,  et~al., 2011, \mn@doi [arXiv e-prints]
  {10.48550/arXiv.1110.3193}, \href
  {https://ui.adsabs.harvard.edu/abs/2011arXiv1110.3193L} {p. arXiv:1110.3193}

\bibitem[\protect\citeauthoryear{{Ledig} et~al.,}{{Ledig} et~al.}{2016}]{SRgan}
{Ledig} C.,  et~al., 2016, \mn@doi [arXiv e-prints]
  {10.48550/arXiv.1609.04802}, \href
  {https://ui.adsabs.harvard.edu/abs/2016arXiv160904802L} {p. arXiv:1609.04802}

\bibitem[\protect\citeauthoryear{{Li}, {Ni}, {Croft}, {Di Matteo}, {Bird}  \&
  {Feng}}{{Li} et~al.}{2021}]{AI1}
{Li} Y.,  {Ni} Y.,  {Croft} R. A.~C.,  {Di Matteo} T.,  {Bird} S.,   {Feng} Y.,
   2021, \mn@doi [Proceedings of the National Academy of Science]
  {10.1073/pnas.2022038118}, \href
  {https://ui.adsabs.harvard.edu/abs/2021PNAS..11822038L} {118, e2022038118}

\bibitem[\protect\citeauthoryear{Maksimova, Garrison, Eisenstein, Hadzhiyska,
  Bose  \& Satterthwaite}{Maksimova et~al.}{2021}]{abacus}
Maksimova N.~A.,  Garrison L.~H.,  Eisenstein D.~J.,  Hadzhiyska B.,  Bose S.,
   Satterthwaite T.~P.,  2021, \mn@doi [Monthly Notices of the Royal
  Astronomical Society] {10.1093/mnras/stab2484}, 508, 4017

\bibitem[\protect\citeauthoryear{{Modi}, {Feng}  \& {Seljak}}{{Modi}
  et~al.}{2018}]{Modi2018}
{Modi} C.,  {Feng} Y.,   {Seljak} U.,  2018, \mn@doi [\jcap]
  {10.1088/1475-7516/2018/10/028}, \href
  {https://ui.adsabs.harvard.edu/abs/2018JCAP...10..028M} {2018, 028}

\bibitem[\protect\citeauthoryear{{Mudur} \& {Finkbeiner}}{{Mudur} \&
  {Finkbeiner}}{2022}]{denoising}
{Mudur} N.,  {Finkbeiner} D.~P.,  2022, \mn@doi [arXiv e-prints]
  {10.48550/arXiv.2211.12444}, \href
  {https://ui.adsabs.harvard.edu/abs/2022arXiv221112444M} {p. arXiv:2211.12444}

\bibitem[\protect\citeauthoryear{{Ni}, {Li}, {Lachance}, {Croft}, {Di Matteo},
  {Bird}  \& {Feng}}{{Ni} et~al.}{2021}]{AI2}
{Ni} Y.,  {Li} Y.,  {Lachance} P.,  {Croft} R. A.~C.,  {Di Matteo} T.,  {Bird}
  S.,   {Feng} Y.,  2021, \mn@doi [\mnras] {10.1093/mnras/stab2113}, \href
  {https://ui.adsabs.harvard.edu/abs/2021MNRAS.507.1021N} {507, 1021}

\bibitem[\protect\citeauthoryear{{Paszke} et~al.,}{{Paszke}
  et~al.}{2019}]{pytorch}
{Paszke} A.,  et~al., 2019, \mn@doi [arXiv e-prints]
  {10.48550/arXiv.1912.01703}, \href
  {https://ui.adsabs.harvard.edu/abs/2019arXiv191201703P} {p. arXiv:1912.01703}

\bibitem[\protect\citeauthoryear{{Perraudin}, {Srivastava}, {Lucchi},
  {Kacprzak}, {Hofmann}  \& {R{\'e}fr{\'e}gier}}{{Perraudin}
  et~al.}{2019}]{Perraudin2019}
{Perraudin} N.,  {Srivastava} A.,  {Lucchi} A.,  {Kacprzak} T.,  {Hofmann} T.,
   {R{\'e}fr{\'e}gier} A.,  2019, \mn@doi [Computational Astrophysics and
  Cosmology] {10.1186/s40668-019-0032-1}, \href
  {https://ui.adsabs.harvard.edu/abs/2019ComAC...6....5P} {6, 5}

\bibitem[\protect\citeauthoryear{{Rodr{\'\i}guez}, {Kacprzak}, {Lucchi},
  {Amara}, {Sgier}, {Fluri}, {Hofmann}  \&
  {R{\'e}fr{\'e}gier}}{{Rodr{\'\i}guez} et~al.}{2018}]{Rodrguez2018}
{Rodr{\'\i}guez} A.~C.,  {Kacprzak} T.,  {Lucchi} A.,  {Amara} A.,  {Sgier} R.,
   {Fluri} J.,  {Hofmann} T.,   {R{\'e}fr{\'e}gier} A.,  2018, \mn@doi
  [Computational Astrophysics and Cosmology] {10.1186/s40668-018-0026-4}, \href
  {https://ui.adsabs.harvard.edu/abs/2018ComAC...5....4R} {5, 4}

\bibitem[\protect\citeauthoryear{{Shi}, {Caballero}, {Husz{\'a}r}, {Totz},
  {Aitken}, {Bishop}, {Rueckert}  \& {Wang}}{{Shi} et~al.}{2016}]{shi2016real}
{Shi} W.,  {Caballero} J.,  {Husz{\'a}r} F.,  {Totz} J.,  {Aitken} A.~P.,
  {Bishop} R.,  {Rueckert} D.,   {Wang} Z.,  2016, \mn@doi [arXiv e-prints]
  {10.48550/arXiv.1609.05158}, \href
  {https://ui.adsabs.harvard.edu/abs/2016arXiv160905158S} {p. arXiv:1609.05158}

\bibitem[\protect\citeauthoryear{{Smith} \& {Lang}}{{Smith} \&
  {Lang}}{2019}]{ytree}
{Smith} B.,  {Lang} M.,  2019, \mn@doi [The Journal of Open Source Software]
  {10.21105/joss.01881}, \href
  {https://ui.adsabs.harvard.edu/abs/2019JOSS....4.1881S} {4, 1881}

\bibitem[\protect\citeauthoryear{{Srisawat} et~al.,}{{Srisawat}
  et~al.}{2013}]{sussing}
{Srisawat} C.,  et~al., 2013, \mn@doi [\mnras] {10.1093/mnras/stt1545}, \href
  {https://ui.adsabs.harvard.edu/abs/2013MNRAS.436..150S} {436, 150}

\bibitem[\protect\citeauthoryear{{Tr{\"o}ster}, {Ferguson},
  {Harnois-D{\'e}raps}  \& {McCarthy}}{{Tr{\"o}ster}
  et~al.}{2019}]{Troster2019}
{Tr{\"o}ster} T.,  {Ferguson} C.,  {Harnois-D{\'e}raps} J.,   {McCarthy} I.~G.,
   2019, \mn@doi [\mnras] {10.1093/mnrasl/slz075}, \href
  {https://ui.adsabs.harvard.edu/abs/2019MNRAS.487L..24T} {487, L24}

\bibitem[\protect\citeauthoryear{{Vogelsberger}, {Marinacci}, {Torrey}  \&
  {Puchwein}}{{Vogelsberger} et~al.}{2020}]{Vogelsberger}
{Vogelsberger} M.,  {Marinacci} F.,  {Torrey} P.,   {Puchwein} E.,  2020,
  \mn@doi [Nature Reviews Physics] {10.1038/s42254-019-0127-2}, \href
  {https://ui.adsabs.harvard.edu/abs/2020NatRP...2...42V} {2, 42}

\bibitem[\protect\citeauthoryear{{Wadekar}, {Villaescusa-Navarro}, {Ho}  \&
  {Perreault-Levasseur}}{{Wadekar} et~al.}{2020}]{wadekar2020hinet}
{Wadekar} D.,  {Villaescusa-Navarro} F.,  {Ho} S.,   {Perreault-Levasseur} L.,
  2020, arXiv e-prints, \href
  {https://ui.adsabs.harvard.edu/abs/2020arXiv200710340W} {p. arXiv:2007.10340}

\bibitem[\protect\citeauthoryear{{Wang}, {Chen}  \& {Hoi}}{{Wang}
  et~al.}{2019}]{srreview}
{Wang} Z.,  {Chen} J.,   {Hoi} S. C.~H.,  2019, \mn@doi [arXiv e-prints]
  {10.48550/arXiv.1902.06068}, \href
  {https://ui.adsabs.harvard.edu/abs/2019arXiv190206068W} {p. arXiv:1902.06068}

\bibitem[\protect\citeauthoryear{{Yao}, {Ishak}, {Lin}  \& {Troxel}}{{Yao}
  et~al.}{2017}]{LSST}
{Yao} J.,  {Ishak} M.,  {Lin} W.,   {Troxel} M.,  2017, \mn@doi [\jcap]
  {10.1088/1475-7516/2017/10/056}, \href
  {https://ui.adsabs.harvard.edu/abs/2017JCAP...10..056Y} {2017, 056}

\bibitem[\protect\citeauthoryear{Yue, Shen, Li, Yuan, Zhang  \& Zhang}{Yue
  et~al.}{2016}]{Yue2016}
Yue L.,  Shen H.,  Li J.,  Yuan Q.,  Zhang H.,   Zhang L.,  2016, \mn@doi
  [Signal Processing] {10.1016/j.sigpro.2016.05.002}, 128

\bibitem[\protect\citeauthoryear{{Zhang}, {Wang}, {Zhang}, {Sun}, {He},
  {Contardo}, {Villaescusa-Navarro}  \& {Ho}}{{Zhang} et~al.}{2019}]{zhang2019}
{Zhang} X.,  {Wang} Y.,  {Zhang} W.,  {Sun} Y.,  {He} S.,  {Contardo} G.,
  {Villaescusa-Navarro} F.,   {Ho} S.,  2019, arXiv e-prints, \href
  {https://ui.adsabs.harvard.edu/abs/2019arXiv190205965Z} {p. arXiv:1902.05965}

\makeatother
\end{thebibliography}

\end{document}